\documentclass[12pt,fleqn,twocolumn]{aastex631}
\usepackage[T1]{fontenc}
\usepackage{ae,aecompl}
\usepackage{graphicx}   % Including figure files

\usepackage{ulem}
\usepackage{amsmath}    % Advanced maths commands
\usepackage{amssymb,bm}    % Extra maths symbols
\usepackage{mathtools}
\usepackage{relsize}
\usepackage{natbib}
\usepackage{hyperref}
\usepackage{color}      
\definecolor{dgreen}{rgb}{0,0.5,0}
\definecolor{magen}{rgb}{0.79,0.08,0.48}
\definecolor{darkred}{rgb}{0.65,0.06,0.37}
\usepackage{booktabs}
\usepackage{multirow}
\usepackage{makecell}

\newcommand{\new}[1]{{\textcolor{black} {#1}}}
\newcommand{\neww}[1]{{\textcolor{black} {#1}}}

\shorttitle{Lensing degeneracies in clusters}
\shortauthors{Williams et al.}

\begin{document}

\title{Shape Degeneracies: the likely culprit for the differences between lens mass models of galaxy clusters}
\author[0000-0002-6039-8706]{Liliya L.R. Williams}
\author[0000-0002-4693-0700]{Derek C. Perera}
\affiliation{Minnesota Institute for Astrophysics, University of Minnesota, 116 Church Street SE, Minneapolis, MN 55455, USA}
\author[0000-0002-3648-8031]{Jori Liesenborgs}
\affiliation{UHasselt – Flanders Make, Digital Future Lab, Wetenschapspark 2, B-3590, Diepenbeek, Belgium}
\author[0000-0002-7876-4321]{Ashish K. Meena}
\affiliation{Indian Institute of Science Bangalore: Bengaluru, Karnataka, India}
\author[0000-0001-6636-4999]{Marceau Limousin}
\affiliation{Aix Marseille Univ, CNRS, CNES, LAM, Marseille, France}
\author[0009-0005-3508-2469]{Leon R. Ecker}
\affiliation{Universit\"ats-Sternwarte M\"unchen, Fakult\"at f\"ur Physik,
Ludwig-Maximilians-Universit\"at M\"unchen,
Scheinerstr.~1, 81679 M\"unchen, Germany}
\affiliation{Max Planck Institute for Extraterrestrial Physics,
Giessenbachstr.~1, 85748 Garching, Germany}
\affiliation{Max-Planck-Institut f\"ur Astrophysik,
Karl-Schwarzschild-Str.~1, 85748 Garching, Germany}

\begin{abstract}
Sky-projected mass distributions of galaxy clusters are widely used to constrain particle properties of dark matter, to magnify in flux and angular extent high redshift background galaxies, and to construct their luminosity function down to faint absolute magnitudes that are otherwise hard or impossible to reach. Obtaining accurate and precise mass maps of clusters is therefore of central importance for these studies. Yet, a visual inspection of any two mass maps of the same cluster shows that the shapes of their isodensity contours are often dissimilar in detail on $\lesssim 30\,$kpc scales, and there is no simple relation connecting the two. These differences are due to systematic uncertainties in reconstructions, which are in turn due to lensing degeneracies. However, commonly known degeneracies, like the mass sheet degeneracy or the source position transformation cannot account for most of these differences, implying that more generalized shape degeneracies are responsible. In this paper we study shape degeneracies using an analytical approach. We conclude with three main takeaways: (i) Shape degeneracies are the likely cause of differences in the mass maps of cluster lenses recovered by different techniques. (ii) Shape degeneracies are significantly suppressed in regions of the lens plane that have tight groupings of lensed images. (iii) \new{In clusters where image numbers are high, such as MACS J0416,} their effect on time delays between images from the same source are generally small, of order $1\%$, but can reach $\sim10\%$ for some multiply imaged background sources, affecting the determination of $H_0$. \\
\end{abstract}

\section{Introduction}

Galaxy clusters are of considerable interest because of their unique ability to tell us about the nature of dark matter, and to magnify high redshift background objects that are mostly out of reach with ground- and space-based telescopes. Clusters have been used to place constraints on the nature of dark matter \citep{diego2024,broadhurst2025}, constrain cosmology \citep{kelly2023,pascale2025,suyu2026}, identify and study individual stars at cosmological distances \citep{kelly18,kelly22,meena22,chen22,diego2023,meena23,furtak2023,golubchik26,hwilliams2026A,hwilliams2026B,diego2026}, look for Pop III stellar populations \citep{fujimoto2025}, place constraints on the high redshift Initial Mass Function and star formation history \cite{li2025A,li2025B,zackrisson2026,chemerynska2026}, characterize small scale subhalos \citep{Perera2026}, etc.

All these goals require accurate and precise maps of cluster mass distributions, which are obtained using gravitational lensing reconstructions. Lensing mass inversion is a mature field, having been around for a couple of decades \citep{natarajan24}. Its significance in astronomy has been highlighted by legacy observational campaigns like 
Cluster Lensing and Supernova Survey with Hubble \citep[CLASH;][]{Postman2012}
Hubble Frontier Fields \citep[HFF;][]{Lotz2017},
Beyond Ultra-deep Frontier Fields And Legacy Observations \citep[BUFFALO;][]{Steinhardt2020},
Prime Extragalactic Areas for Reionization and Lensing Science \citep[PEARLS;][]{Windhorst2023},
Strong Lensing and Cluster Evolution \citep[SLICE;][]{Mahler2024}, 
Vast Exploration for Nascent, Unexplored Sources \citep[VENUS;][]{fujimoto2025}, 
JWST Advanced Deep Extragalactic Survey \citep[JADES;][]{Eisenstein2026}), etc. Several independent lens inversion methods are used to convert these data into mass models of clusters \citep[see][]{natarajan24}. A handful of prominent cluster-lenses, with more than 100 or so lensed images have been reconstructed by several groups, often more than once.  

An outsider may wonder why a cluster constrained by a large number of lensed images needs several modeling teams to reconstruct its mass. The reason is that different modeling philosophies, or different priors with the same modeling code usually result in somewhat different maps, and it is not clear which of these, if any, are closest to the truth. While all methods largely agree on global properties, like the total mass, clusters' ellipticity and its position angle, and major substructures, the differences are mostly in the details, like smaller scale features. In the era of precision cosmology, even small differences can be significant, for example, for the determination of the luminosity function of high redshift sources \citep{bouwens2017,atek2018,bouwens2022,atek2026}.

Given the high level of interest in clusters as gravitational lenses, several studies have carefully examined the resulting mass reconstructions. Specifically, a few meta-studies comparing results of various reconstructions exist \citep{meneghetti2017,priewe2017,remolina2018,raney2020b}. The emphasis and the details of the findings differ between studies, but all indicate that models differ from each other, often by more than their stated statistical uncertainties. 

%on galaxy scales, there are also studies comparing models:
%\cite{galan2024,ding2021}.

At this point there is about a decade worth of accumulated models for a few clusters, so one can assess the progress that the cluster inversion community has made since the launch of HFF.  The most recent meta-study did just that. \cite{Perera2025} asked if the models' recovered mass distributions have been converging or diverging as the number of multiple images increased. The hope is that with the increasing number of lensed images, the models will converge to each other and, by induction, to the true mass distribution.
Surprisingly, the paper showed that if one compares any two lens models by subtracting their 2D spatial mass maps and taking the median of the differences over the cluster region that contains images, there is, typically, a $\sim9.0\%$ difference between any two models. This applies to differences between reconstructions using different methods---simply parametrized vs. free-form---but the typical differences between two simply parametrized models are equally large. 

Parametrized and free-form models employ very different approaches: the former represents cluster mass as a superposition of simple elliptical cluster-wide potentials (these usually number in the low single digits), and individual cluster galaxies, whose mass is usually related to their observed luminosity or velocity dispersion. These simple parameterizations capture the main mass features one expects to find in a cluster. But clusters, especially merging ones, are unlikely to be accurately represented by such idealized mass distributions, and probably host more messy signatures of past and ongoing mergers.  Free-form models, which use more general mass basis functions were developed to address these. Hybrid models include features of both types of methods, making sure that the invisible dark matter is not too constrained by restrictive parametric mass forms.

Part of the reason for differences between mass maps of different methods is that not all reproduce image positions equally well. But this is not the only factor, and possibly not even the main one because lens plane image rms---the difference between the observed and reconstructed image positions in the observer's frame---is quite similar for most methods for the same cluster, and is generally of order of an arcsecond or smaller. The other reason, possibly the dominant one, is the lensing degeneracies. Model degeneracies are encountered in many sub-fields of astrophysics, and lensing is no exception.

The analysis presented in \cite{Lasko2023} suggests that multiple images in clusters do not pinpoint substructure well; each model's reconstructed substructure is in part determined by the prior assumptions used, like the mass basis set and parameter ranges. This leads to different models recovering somewhat different mass maps, related by degeneracies.

Mass Sheet Degeneracy \citep[MSD;][]{falco1985,saha2000}, is widely recognized in all lensing work. Exact MSD, which can be broken by sources at more than one redshift has been shown to be generalizable to cases with multiple source redshifts by modifying the mass distribution entirely within the main lens plane \citep{liesenborgs2008a,liesenborgs2012}, or by introducing appropriately scaled mass sheets at various source planes \citep{schneider2014b,schneider2019,teodori2026}, making clusters susceptible to these. Monopole degeneracy can redistribute mass in any circularly symmetric fashion in regions outside of the observed image locations \citep{saha2000,liesenborgs2008b,liesenborgs2012,liesenborgs2024}, and 
source position transformation \citep[SPT; ][]{schneider2014a,unruh2017} alters the deflection field and source positions, while preserving image positions. \neww{In the context of the parametrized lens inversion methods, there is another type of degeneracy that is relevant, between the mass in the smooth cluster-scale dark matter component and the cluster galaxies component, which can hinder further insights into the dark matter properties \citep[e.g.][]{limousin2016}. Even if a major improvement has been achieved using spectroscopic observations of cluster members in order to anchor their mass \citep{bergamini2019}, this degeneracy still persists, as has been recently shown in \cite{limousin2025b,Limousin2025}.}

This paper explores a wide class of lensing degeneracies, those that do not have well defined contours of the mass distribution, but are instead amorphous in shape. It is apparent that shape degeneracies are common: a simple visual inspection of any two mass reconstructions of the same cluster shows that the shapes of the density contours are different, and there is no simple relation that connects the two \citep[e.g.,][]{Bergamini2023,rihtarsic2025}. That means any two reconstructions are related by shape degeneracies.
Shape degeneracies were first introduced by \cite{Saha2006} in the context of galaxy-scale lenses, though that paper's description was qualitative, and did not present a quantitative way to construct these.\footnote{Examples of galaxy-scale shape degeneracies in recent parametric reconstructions can be found in \cite{miller2025}.}

From the point of view of lensing, shape degeneracies are low amplitude mass features that do not have well defined shapes or isodensity contours in the lens plane. From the astrophysical point of view, clusters can have low amplitude amorphous mass features as a consequence of mergers and interactions within clusters, \neww{similar to dark matter tidal features found in simulations of the Local Group \citep{garavito2019,tamfal2021}.} The resulting mass features need not obey simple parametric forms. Spatial scales in clusters are large, and densities in the regions between galaxy members can be low enough to lead to long dynamical and relaxation timescales. This means that even in approximately equilibrium clusters, non-equilibrium, low amplitude mass features can persist for a long time.

One specific application of the analysis we present concerns M2, a $M\approx {\rm few}\,10^{11}M_\odot$ mass clump recovered by \cite{perera2024b} in galaxy cluster MACS J0416, and examined in \cite{Limousin2025}, using different modeling methods. No two models fully agreed with each other on the existence, shape and mass of M2. Most parametric models did not recover it at all. One of the goals of this paper is to determine if shape degeneracies can make such mass clumps appear in some models, and be absent on others.

For brevity, we call shape degeneracies ShaDes, and define them as transformations that preserve the deflection angles at the locations of all the observed images exactly. 
Given the lens equation, $\bm{\theta}=\bm{\beta}+\bm{\alpha}(\bm{\theta})$, to preserve deflection angles means that the source positions also need to be the same as in the original model. 
In Section~\ref{sec:constructing} we show how these can be constructed using linear algebra. In Section~\ref{sec:macs0416} we apply ShaDes to a galaxy cluster MACS J0416. In Section~\ref{sec:tight} we examine the role of tight groupings of images in suppressing ShaDes. In Section~\ref{sec:m2} we examine their effect on the mass clump M2, and in Section~\ref{sec:h0} we look into their effects on the determination of the Hubble constant.
%In Section~\ref{sec:beyond} we use the linear algebra formalism employed in ShaDes to go beyond them. 
In Section~\ref{sec:offset} we use a modified version of the lens equation where the sources in the source plane are offset, and the corresponding image deflection angles compensate for that exactly, leaving image positions unchanged. 

Until this point in the paper we assume that model reconstructed image positions are exactly the same for all models. This is not the case: there is a typical difference of  $\lesssim 1''$ between observed and reconstructed images, and between images of different models. In Section~\ref{sec:shades+} we incorporate that fact into our analysis and build approximately degenerate solutions; we call these ShaDes+. A discussion of how shape degeneracies fit into the broader landscape of lensing degeneracies is in Section~\ref{sec:disc}, and we summarize in Section~\ref{sec:conc}.

\section{Constructing S\lowercase{ha}D\lowercase{es}}\label{sec:constructing}

We construct degenerate mass models of clusters as a linear superposition, $\mathcal{S}(\bm{\theta})$, of an existing 2D mass model, $\mathcal{M}(\bm{\theta})$, from any modeling method, and a 2D density perturbation field, $\mathcal{P}(\bm{\theta})$, such that, $\mathcal{S}(\bm{\theta})=\mathcal{M}(\bm{\theta})+\mathcal{P}(\bm{\theta})$, where $\bm{\theta}$ is the angular location in the cluster. While the existing models necessarily have $\kappa>0$ everywhere, the perturbation field can have positive and negative $\Delta\kappa$ values at different locations in the cluster. $\kappa$ is the projected surface mass density, normalized by critical density for lensing.\footnote{That critical density is $\Sigma_{\rm crit}=\frac{c^2}{4\pi G} \frac{D_{os}}{D_{ol}D_{ls}}$, and depends on the lens and source redshifts through that combination of angular diameter distances. In this work we set $D_{os}/D_{ls}=1$, removing the dependence on source redshifts.} We require that the amplitude of perturbations is such that the overall $\kappa>0$ over the whole cluster. 

The same existing model, $\mathcal{M}(\bm{\theta})$, combined with two different perturbation fields, $\mathcal{P}_1(\bm{\theta})$ and $\mathcal{P}_2(\bm{\theta})$, leads to two degenerate mass models,
$\mathcal{S}_1(\bm{\theta})=\mathcal{M}(\bm{\theta})+\mathcal{P}_1(\bm{\theta})$ and $\mathcal{S}_2(\bm{\theta})=\mathcal{M}(\bm{\theta})+\mathcal{P}_2(\bm{\theta})$. Any number of such degenerate models can be constructed. Because deflection angles from various mass elements add up linearly, we use linear algebra to construct perturbation fields, $\mathcal{P}_i(\bm{\theta})$. In general the resulting deflection angle at an arbitrary location $\bm{\theta}$ in the lens plane is
\begin{equation}
{\bm{\alpha}}_{\mathcal{P}}({\bm{\theta}}) = \sum_{j=1}^{2N} {b_j} {\bm{\alpha}}_{{\rm basis},j}(\bm{\theta}).
\end{equation}
%Here, $\alpha_{x,i}$ are the $x$-deflection angles for the $i$th image, and there are a total of $N$ lensed images. 
The amplitudes, or weights of the basis functions are given by $b_j$, %their locations in the lens plane are ${\bm{\theta}}_j$ 
and their deflection angles are ${\bm{\alpha}}_{{\rm basis},j}({\bm{\theta}})$ at $\bm\theta$.
%Each equation sums up the contributions of all basis functions at each image. 
Because, by definition, degenerate mass models must reproduce image positions as well as the original mass model, the deflection angles from any given $\mathcal{P}_i(\bm{\theta})$ at all image positions must be zero. In other words, $\mathcal{P}_i(\bm{\theta})$ must leave image positions unchanged, which means that
\begin{equation}
{\bm{\alpha}}_{\mathcal{P}}({{\bm{\theta}_{\rm im}}}) = \sum_{j=1}^{2N} {b_j} {\bm{\alpha}}_{{\rm basis},j}({\bm{\theta}}_{\rm im})=0,
\end{equation}\label{eq:N0Ds}
where ${\bm{\theta}}_{\rm im}$ are the positions of the observed images on the plane of the sky.\footnote{\new{In principle, ShaDes can be applied either to the observed image positions, leaving these unchanged, or to the predicted images of any given lens model. Here we use the former approach.}}
Since there are $N$ equation for each of $x$- and $y$-deflections,  there are $2N$ equations in all, so we are allowed to have $2N$ basis functions. \new{More generally, there can be more basis functions, but $2N$ are uniquely determined, and this is what we will be using here.}

We aim to have a flexible basis set that can represent a wide variety of $\mathcal{P}_i(\bm{\theta})$ density perturbation shapes over the face of a galaxy cluster. We choose a basis set of circularly symmetric profiles, {\tt alphapots}, which have simple analytical expressions for convergence ($\kappa$), deflection angles, and lensing potential \citep{keeton2001},
\begin{equation}
   \Psi(\theta)=b(\theta^2+s^2)^{\gamma/2}, 
\end{equation}
where $\theta$ is the 2D projected distance (in angular units) from the center of that {\tt alphapot}, $s$ is the core radius, and $\gamma$ is the slope of the potential, with the projected slope of the corresponding density profile being $\Delta\kappa\propto r^{\gamma-2}$. We use profile slopes that ensure that {\tt alphapot}'s density decreases steeply with distance from its center. Any other choice of circularly symmetric basis functions would work equally well. In the future it may be interesting to try elliptical basis functions. Unlike the basis set of the monopole degeneracy, the total mass in our bases need not add up to zero. We work in the space of deflection angles, which are  first gradients of the lensing potential. The curl-free nature of the total potential is guaranteed because it is the sum of individual {\tt alphapots} that satisfy the curl-free condition.

To generate $\mathcal{P}_i(\bm{\theta})$ we use a set of observed $N$ multiple images, then randomly choose centers of the $2N$ {\tt alphapots}, and solve the set of eqs.~\ref{eq:N0Ds} using LU matrix decomposition. To prevent the inversion from returning all $b_j=0$, we initialize by introducing one or two additional mass clumps, also modeled as {\tt alphapots}, to non-zero values, and keeping these fixed. A given randomly chosen set of {\tt alphapot} positions will produce a solution where deflection angles are zero at the images. However, some of these solutions will have locations in the lens plane where $\Delta\kappa$ due to $\mathcal{P}_i(\bm{\theta})$ has unphysically large absolute values. So we keep only those $\mathcal{P}_i(\bm{\theta})$ that result in the range of $\Delta\kappa$ amplitudes that we desire. Alternatively, any basis set that is already a solution, can be rescaled in amplitude to have higher or lower excursions in $\Delta\kappa$.

\neww{Adding ShaDes to an existing lens model can be compared to adding B-spines to an existing {\rm Lenstool} model, as was done in \cite{beauchesne2021}. However, in the former case, ShaDes do not alter image positions, while in the latter, the added B-splines are meant to improve the correspondence between observed and model predicted image positions.}

Most of our ShaDes analysis treats sources at all redshifts the same way, since ShaDes are applied to the mass of the cluster, not along the line of sight. In Section~\ref{sec:offset} we will allow sources at different redshifts to be offset by various amounts.

In this paper we restrict ourselves to point images. Extended images will likely lead to a suppression of shape degeneracies. There are two effects to consider. First, because extended images fill detector pixels with flux, the  deflection angles at these pixels can be as large as the pixel size, which for JWST, for example, is $0.03''-0.1''$. This by itself will allow somewhat more freedom for shape degeneracies than point sources would. However, the more dominant effect is that extended images cover multiple contiguous pixels, requiring the deflection angles to be preserved, to pixel scale, over extended regions in the source plane. This second effect will suppress shape degeneracies.  If deflection angles from ShaDes are small over extended lens plane regions, that means their gradient is also small, which translates into $\Delta\kappa\approx0$ where extended images exist. Extended sources are somewhat analogous to a tight grouping of point images, which we examine in Section~\ref{sec:tight}.

%Which one of these two effects will dominate depends on the size of the image, and likely other factors as well. We leave this analysis for a future paper. XXX

%%Equations~\ref{eq:N0Ds} assume that all the source positions are exactly the same as in the original unperturbed $\mathcal{M}(\bm{\theta})$ model. This setup can already produce a wide range of degenerate solutions. However, we can also allow some displacement of each source with respect to other sources. This is implemented as a small additive term to $x$- and $y$-deflection angle equations of the images of a given source. Each source (and its images) can have its own additive terms, chosen randomly, i.e., a displacement in the source plane.

\begin{figure*}
    %\vspace{-4cm}
    \centering
    \includegraphics[trim={0.7cm 5cm 1cm 4cm},clip,width=0.32\linewidth]{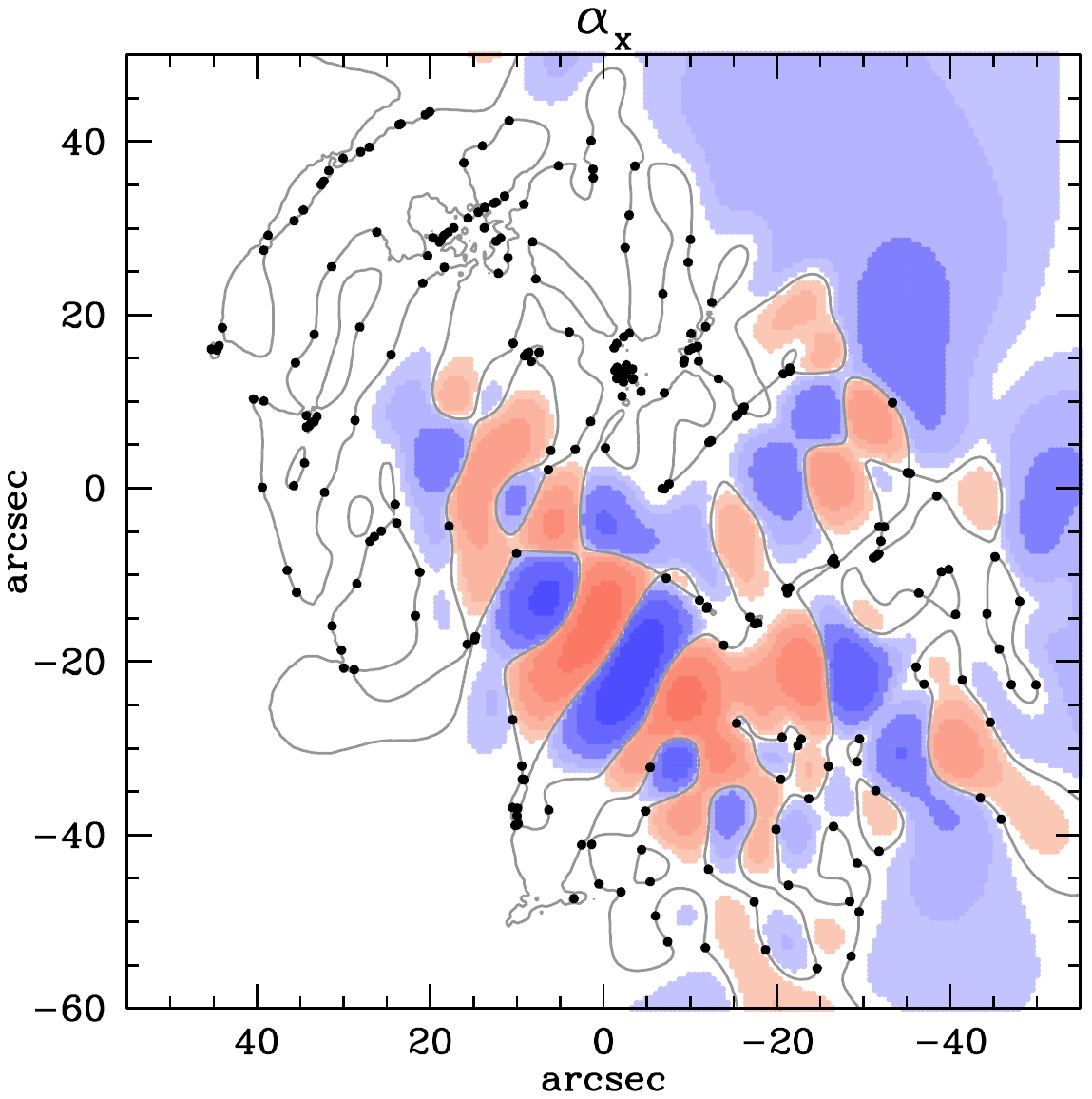}
    \includegraphics[trim={0.7cm 5cm 1cm 4cm},clip,width=0.32\linewidth]{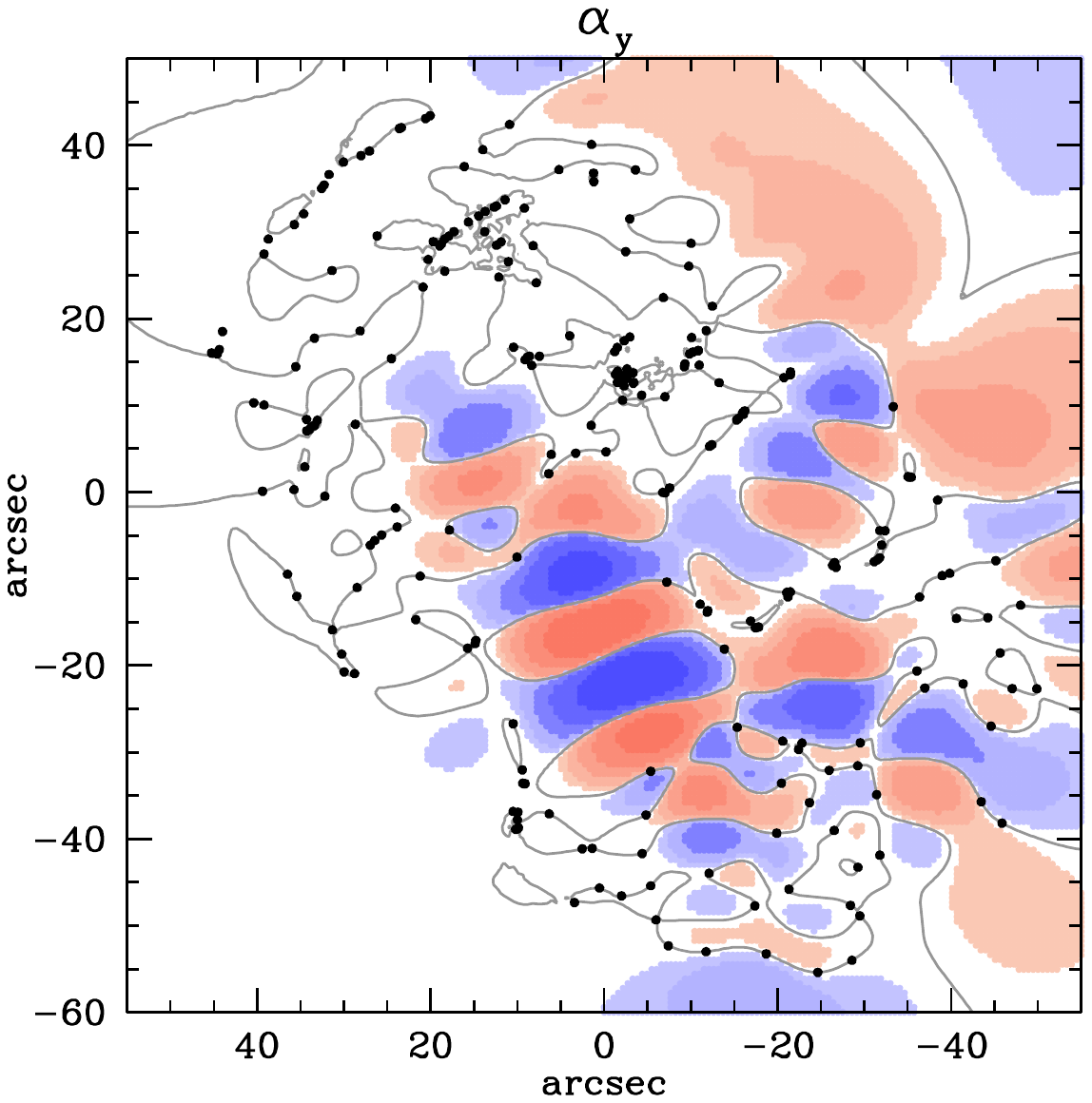}
    \includegraphics[trim={0.7cm 5cm 1cm 4cm},clip,width=0.32\linewidth]{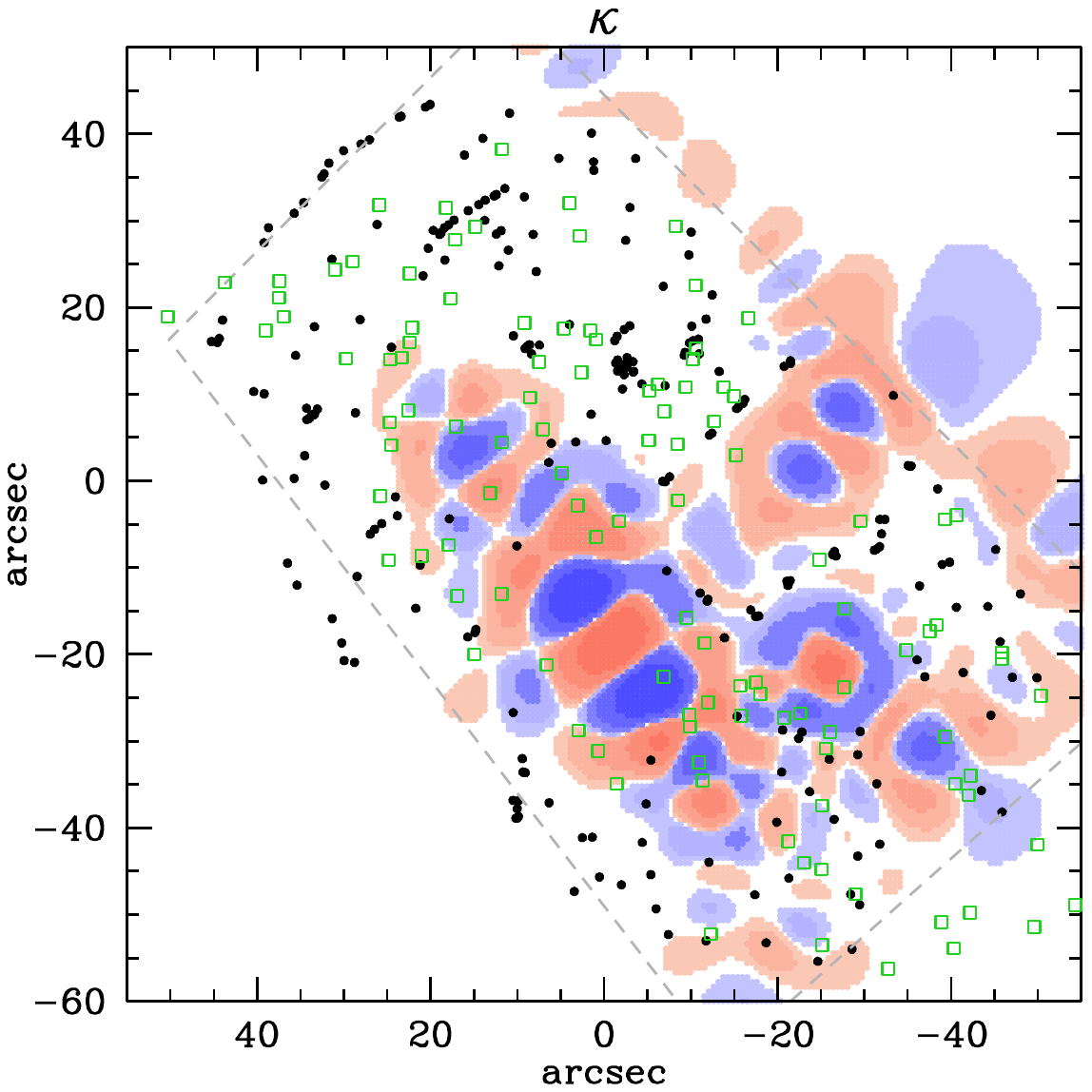}
    %\vspace{-1cm}
    \caption{An example of shape degeneracies, ShaDes, $\mathcal{P}(\bm{\theta})$. Black dots are 237 spectroscopically confirmed images of MACS J0416. Gray lines are contours of zero deflection angles in $x$ ({\it left panel}), and $y$ ({\it middle}). These contours go through all the observed images. Positive and negative deflection angle values are denoted by red and blue colors respectively. \neww{The steps in the color gradation correspond to magnitudes of deflection angles (in arcsec) of $5.5\times 10^{-3}-1.65\times10^{-2}$, $1.65\times 10^{-2}-4.95\times10^{-2}$, $4.95\times 10^{-2}-1.485\times 10^{-1}$, etc. with each interval covering a factor of 3 in deflection angle. }The {\it right panel} shows the $\Delta\kappa$ map of this $\mathcal{P}(\bm{\theta})$. White regions have $|\Delta\kappa|<0.002$. \neww{The grades of red and blue colors show $|\Delta\kappa|$ ranges of $0.002-0.006$, $0.006-0.018$, $0.018-0.054$, etc. with each interval covering a factor of 3 in $|\Delta\kappa|$. (All other figures use the same color scheme.)} Note that $\Delta\kappa$ can be non-zero at the locations of some images. The gray slanted rectangle is the region within which centers of ShaDes basis functions were randomly scattered. Green squares show positions of 119 cluster galaxies; these were not used in creating ShaDes. Values characterizing deflection angles and $\Delta\kappa$ are recorded in Table~\ref{tab:summary}.}
    \label{fig:07F}
\end{figure*}

\begin{figure*}
    \centering
    \includegraphics[trim={0.7cm 5cm 1cm 4cm},clip,width=0.32\linewidth]{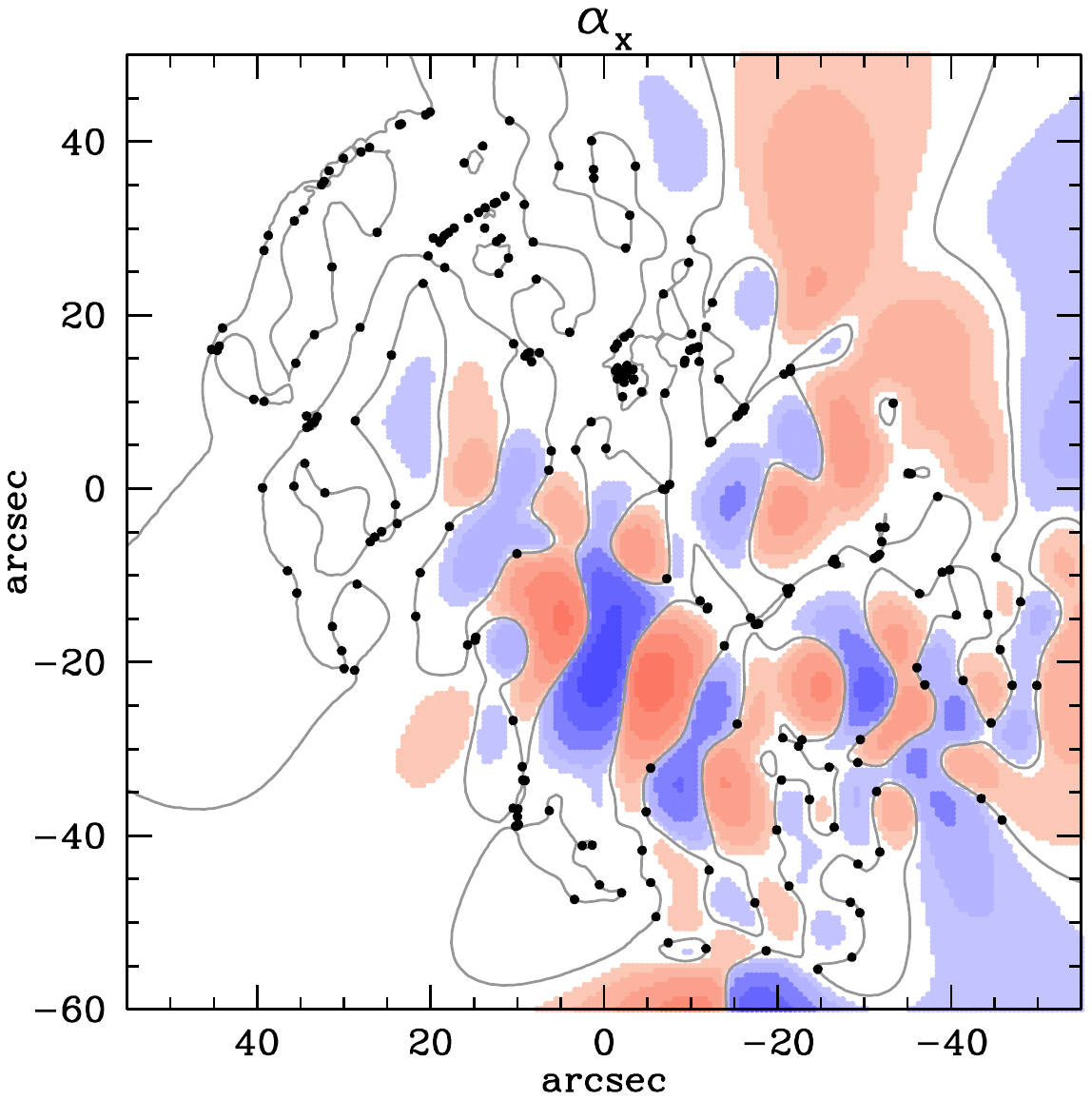}
    \includegraphics[trim={0.7cm 5cm 1cm 4cm},clip,width=0.32\linewidth]{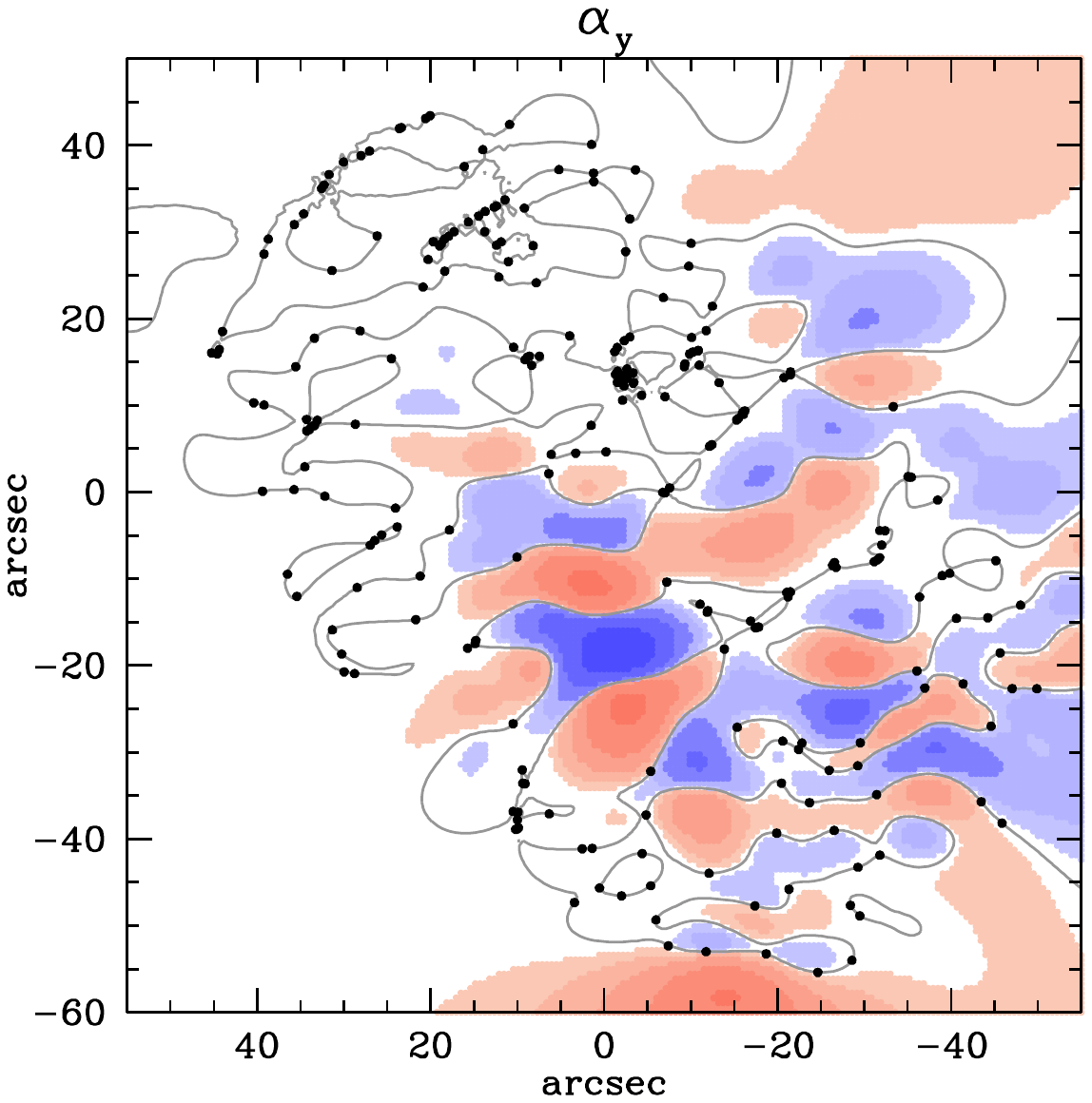}
    \includegraphics[trim={0.7cm 5cm 1cm 4cm},clip,width=0.32\linewidth]{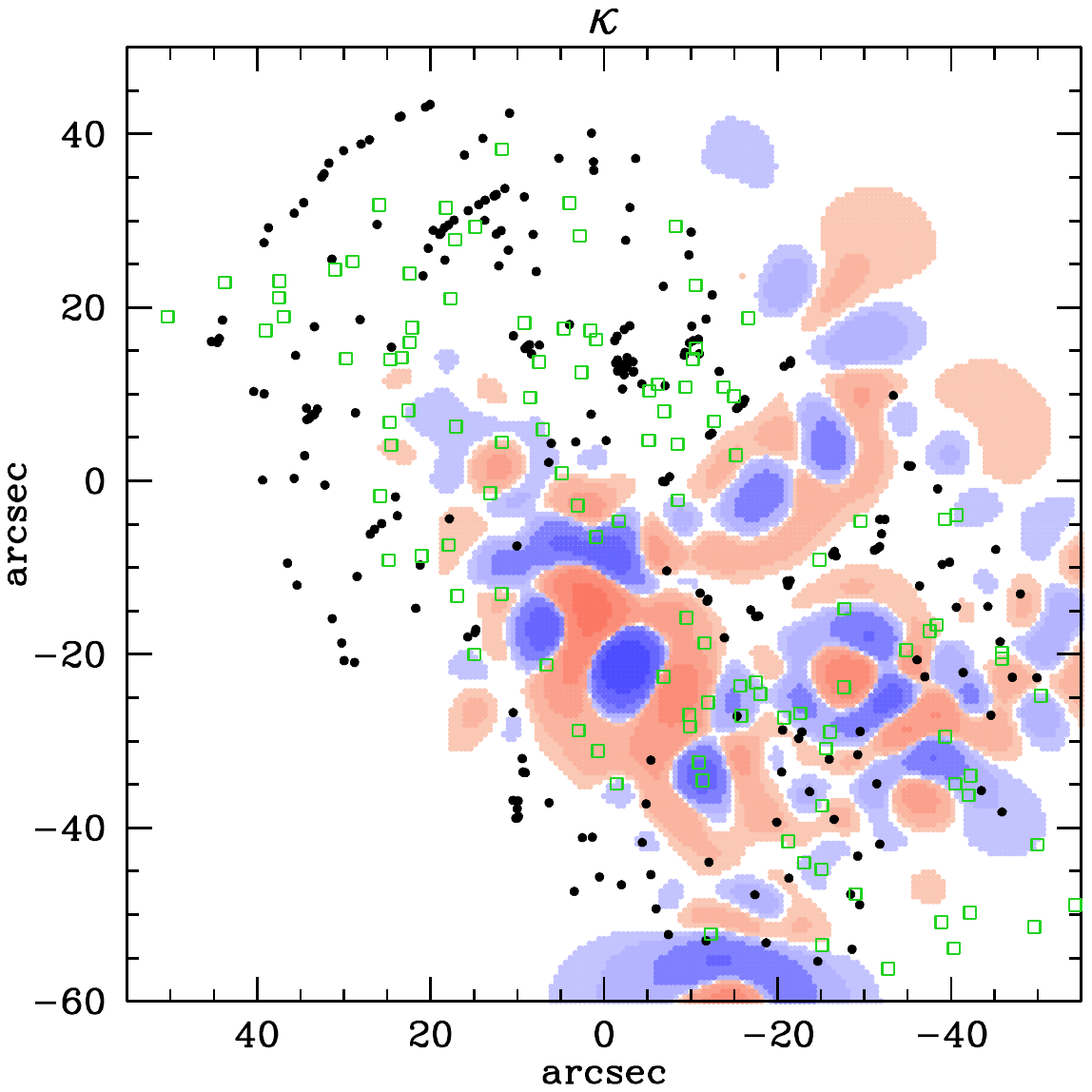}
    %\vspace{-1cm}
    \caption{Similar to Figure~\ref{fig:07F}; see Table~\ref{tab:summary} for values characterizing deflection angles and $\Delta\kappa$.}
    \label{fig:07A}
\end{figure*}

\begin{figure*}
    %\vspace{-4cm}
    \centering
    \includegraphics[trim={0.7cm 5cm 1cm 4cm},clip,width=0.32\linewidth]{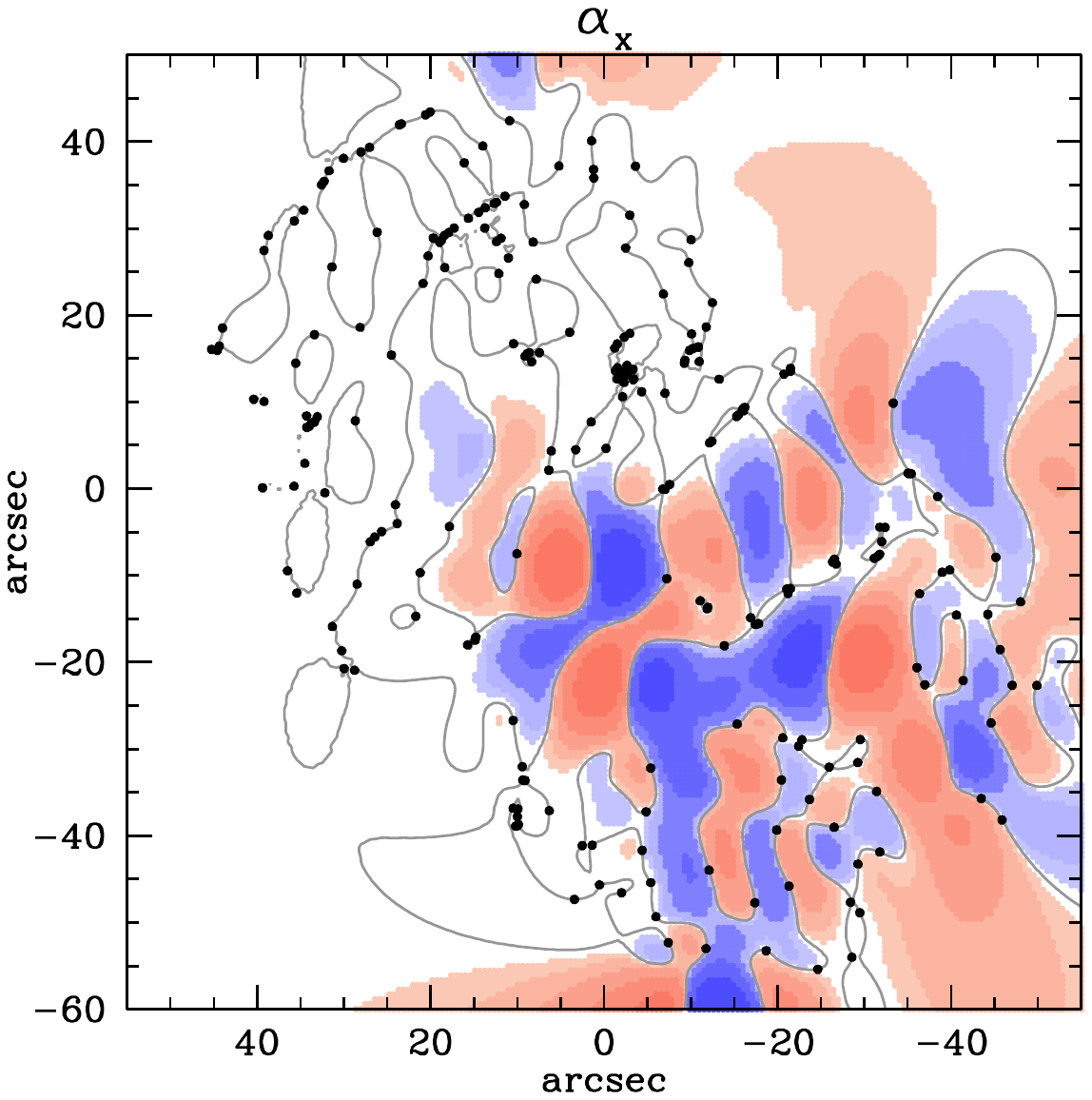}
    \includegraphics[trim={0.7cm 5cm 1cm 4cm},clip,width=0.32\linewidth]{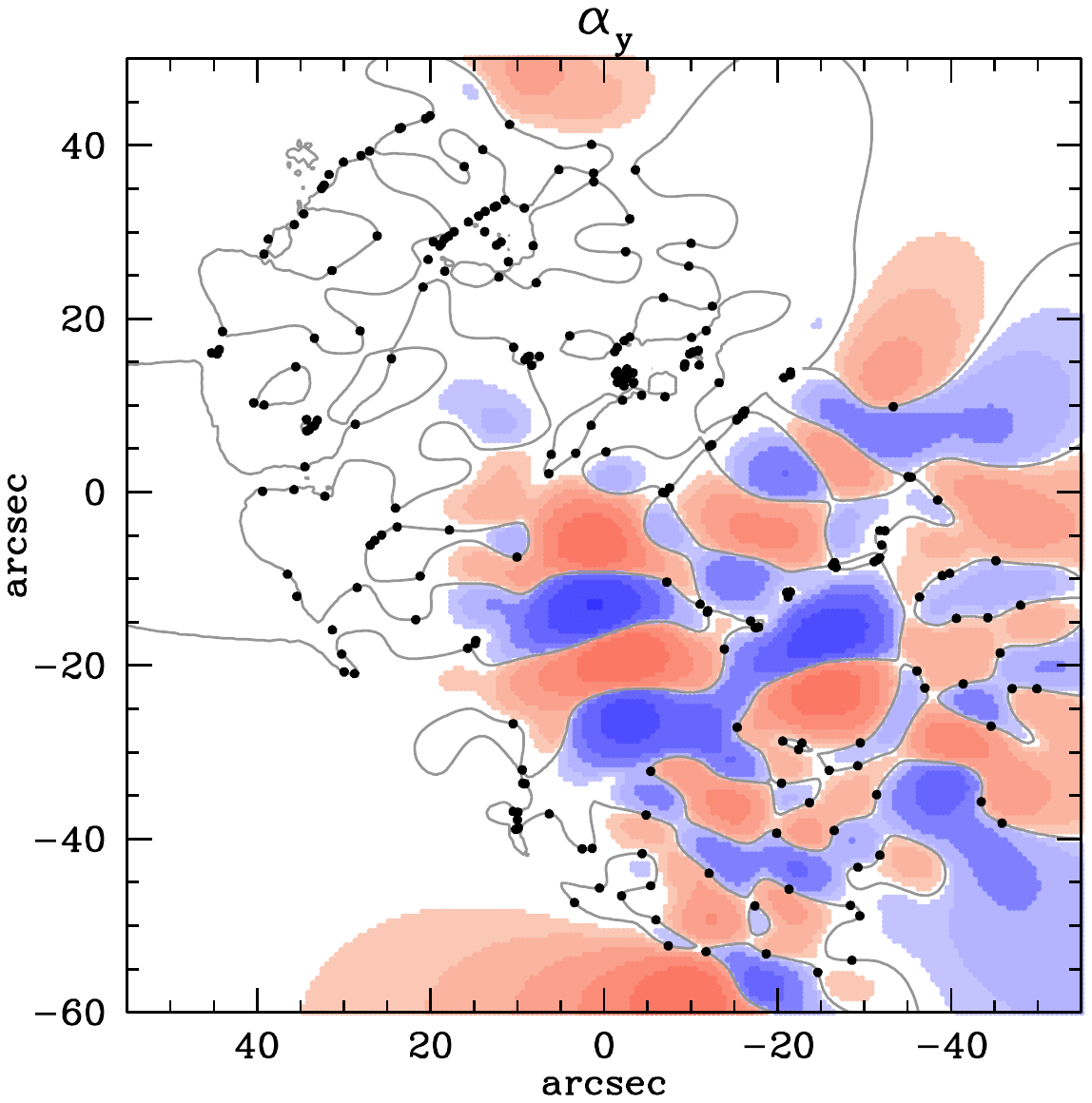}
    \includegraphics[trim={0.7cm 5cm 1cm 4cm},clip,width=0.32\linewidth]{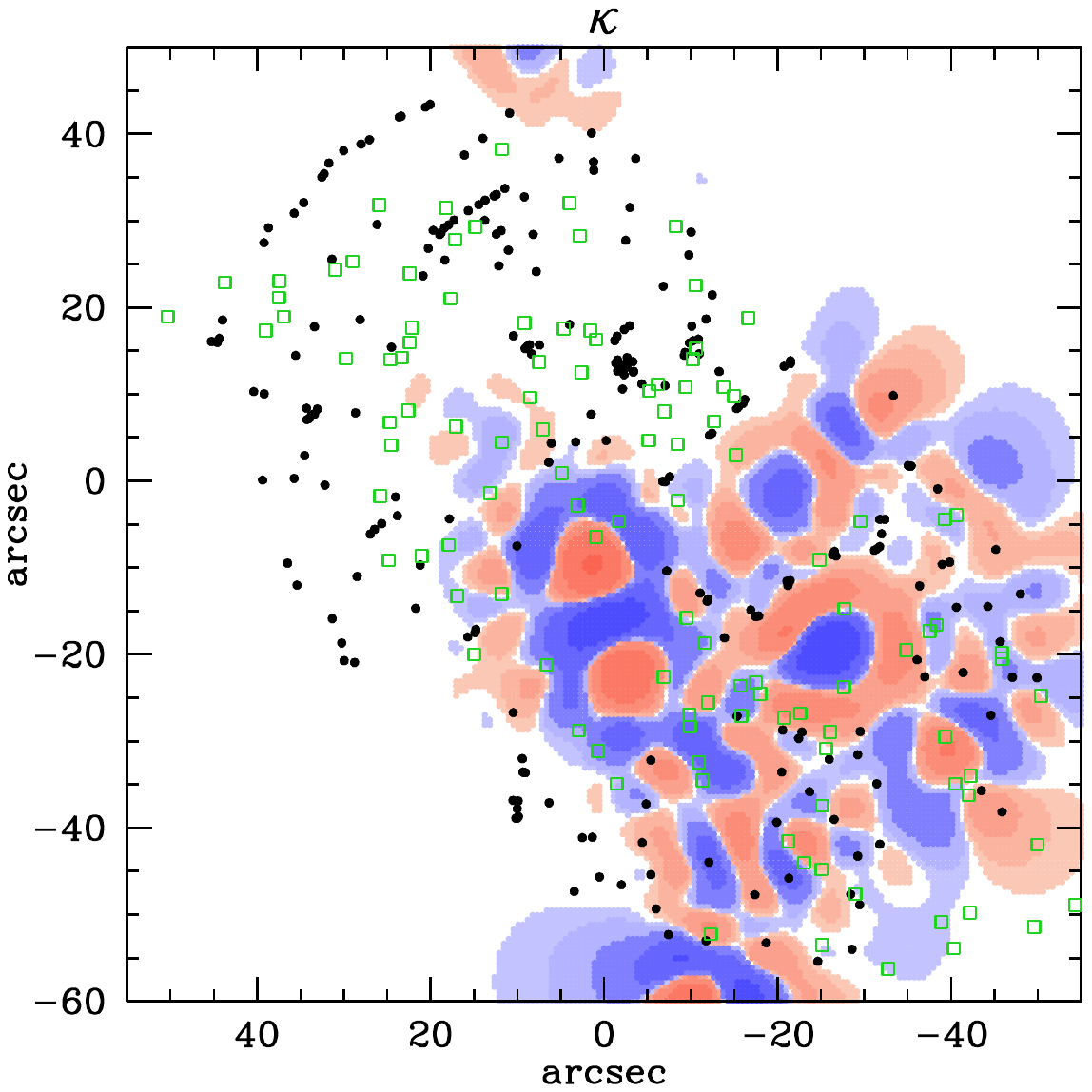}
    %\vspace{-1cm}
    \caption{Similar to Figure~\ref{fig:07F}; see Table~\ref{tab:summary} for values characterizing deflection angles and $\Delta\kappa$.}
    \label{fig:07D}
\end{figure*}

\section{Application to galaxy cluster MACS J0416}\label{sec:macs0416}

MACS J0416, at $z=0.396$, is one of the Hubble Frontier clusters, and has been observed extensively with HST and JWST resulting in $343$ images at a range of redshifts \citep{Bergamini2023,rihtarsic2025}. We use $237$ images from $88$ sources that have been spectroscopically confirmed, and compiled by \cite{Bergamini2023}.

We chose the centers of the basis functions randomly, but confine them to within the slanted rectangle that encloses all the observed images, shown in the right panel of Figure~\ref{fig:07F} with dashed gray lines \citep[Fig.~1 of][]{perera2024b}. Basis functions' core sizes are varied between approximately $5''$ and $15''$; we used a range of $\gamma$ between $0.05$ and $0.25$, all of which result in steep outer density slopes. Fixed initializing clumps have similar parameters, with their positions chosen randomly.

Figures~\ref{fig:07F}-\ref{fig:07D} show examples $\mathcal{P}_i(\bm{\theta})$, as generated by our method. The left and central panels show $x$ and $y$ deflection angles of $\mathcal{P}_i(\bm{\theta})$, with the gray lines outlining contours of zero deflections. These contours go through all the 237 images. Red and blue colors show the positive and negative magnitudes of deflection angles, respectively. The $\Delta\kappa$ maps are in the right panels, where green squares denote the positions of cluster galaxies. These are not used in our analysis, and are shown for reference only. %%(We assumed $D_{ls}/D_{os}=1$.)
The positive (red) and negative (blue) mass clumps that are apparent in the Figures do not correspond to individual basis functions, instead, they are a combination of various bases. 

The extreme values of the deflection angles and $\Delta\kappa$ values for each of the Figures are recorded in  Table~\ref{tab:summary}. 

When any of these $\mathcal{P}_i(\bm{\theta})$ are added to any existing mass reconstruction $\mathcal{M}(\bm{\theta})$ of MACS J0416, simply parametrized, free-form or hybrid, the result is a degenerate model that preserves image positions, $\mathcal{S}_i(\bm{\theta})$. The $\kappa$ values of $\mathcal{S}_i(\bm{\theta})$ do not go below zero in the cluster, though some very small negative values, $\kappa\sim -0.001$ may occur further away from the cluster. This applies to all the models we present in this paper.

The last column of Table~\ref{tab:summary} gives the value of the median percent difference, MPD (see \cite{Perera2025} for the definition) between the $\kappa$ maps of {\tt Grale} and {\tt Grale} with ShaDes added, calculated within the rectangle shown in the right panel of Figure~\ref{fig:07F}. These MPD values are small compared to the typical difference between any two models of the same cluster, which are around 9\%. We return to this in Section~\ref{sec:shades+}.

Though all 3 examples shown in Figures~\ref{fig:07F}-\ref{fig:07D} are different, there are commonalities: the upper left region of the cluster does not have any noticeable deviations from $\Delta\kappa=0$, even though the centers of {\tt alphapots} are distributed randomly within the cluster, and populate that region as well. This is likely because there exist a few tight groupings of images in that region, notably, the Warhol arc around $(-3'',12'')$.  This suggests that the best guard against shape degeneracies is the presence of image groups in several regions of the lens plane. 

\neww{Related to this, we make a note on image magnifications. While ShaDes preserve image positions, image magnifications will not be preserved. Figures~\ref{fig:07F}-\ref{fig:07D} show that typical $\Delta\kappa$ values are not large, but even small changes in surface density can still lead to non-negligible changes in image magnifications \citep{limousin2016,limousin2025b}. This suggests that the regions of the cluster where ShaDes are suppressed (upper left in the case of MACS J0416) will yield the smallest systematic uncertainties on magnification of background sources.} 

The spatial (i.e., angular) scale of $\mathcal{P}_i(\bm{\theta})$ perturbations is set by the typical image separation, and the spatial scale of the basis functions used. Since we can control the latter, we have experimented with using larger and smaller scale basis functions.
If we use much smaller {\tt alphapot} scale by reducing their core size, $s$, or making their density profile steeper, that still does not populate the upper region of the cluster with ShaDes. On the other hand, using larger scale basis functions suppresses the amplitude of shape perturbations. So the only relevant scale is that corresponding to the typical image separation. 

\begin{table*}[]
    \centering
    \begin{tabular}{l|ccc|ccc|ccc|c}
    \hline
Figure & min $\alpha_x$ & max $\alpha_x$ & rms $\alpha_x$ & min $\alpha_y$ & max $\alpha_y$ & rms $\alpha_y$ & min $\Delta\kappa$ & max $\Delta\kappa$ & rms $\Delta\kappa$ & MPD \\
    \hline
Fig.~\ref{fig:07F}          & -0.87 &  0.84 & 0.108 & -1.20 &  1.16 & 0.137 & -0.39 &  0.43 & 0.043  &  0.42 \\  %  07F
Fig.~\ref{fig:07A}          & -0.83 &  0.76 & 0.081 & -0.92 &  0.58 & 0.086 & -0.40 &  0.25 & 0.028  &  0.29 \\  %  07A
Fig.~\ref{fig:07D}          & -1.18 &  1.04 & 0.153 & -1.39 &  1.19 & 0.174 & -0.43 &  0.54 & 0.054  &  0.67 \\  %  07D
Fig.~\ref{fig:1106} (left)  & -0.70 &  0.70 & 0.072 & -0.73 &  0.43 & 0.058 & -0.37 &  0.24 & 0.024  &  0.16 \\  %  11C
Fig.~\ref{fig:1106} (middle)& -0.90 &  1.29 & 0.128 & -0.61 &  0.84 & 0.101 & -0.38 &  0.24 & 0.035  &  1.82 \\  %  06C
Fig.~\ref{fig:1106} (right) & -1.05 &  0.94 & 0.105 & -0.83 &  0.62 & 0.077 & -0.50 &  0.28 & 0.033  &  0.66 \\  %  06D
Fig.~\ref{fig:14} (left)    & -1.89 &  1.78 & 0.097 & -1.57 &  1.50 & 0.080 & -0.15 &  0.24 & 0.020  &  0.17 \\  %  14A
Fig.~\ref{fig:14} (middle)  & -0.33 &  0.36 & 0.022 & -0.38 &  0.34 & 0.024 & -0.03 &  0.04 & 0.005  &  0.08 \\  %  14F
Fig.~\ref{fig:14} (right)   & -1.94 &  1.85 & 0.112 & -1.68 &  1.60 & 0.113 & -0.21 &  0.29 & 0.027  &  0.29 \\  %  14D
Fig.~\ref{fig:08} (left)    & -1.19 &  1.04 & 0.136 & -1.68 &  1.19 & 0.153 & -0.56 &  0.30 & 0.045  &  0.38 \\  %  08E
Fig.~\ref{fig:08} (middle)  & -0.58 &  0.73 & 0.065 & -0.72 &  0.69 & 0.071 & -0.30 &  0.10 & 0.021  &  0.24 \\  %  08B
Fig.~\ref{fig:08} (right)   & -0.66 &  0.53 & 0.068 & -0.97 &  0.74 & 0.088 & -0.34 &  0.22 & 0.025  &  0.28 \\  %  08A
Fig.~\ref{fig:17A}          & -1.60 &  0.93 & 0.283 & -0.90 &  0.89 & 0.205 & -0.27 &  0.24 & 0.053  &  4.05 \\  %  17A
Fig.~\ref{fig:17F}          & -0.68 &  1.22 & 0.240 & -1.05 &  1.28 & 0.283 & -0.15 &  0.15 & 0.038  &  2.80 \\  %  17F
%Fig.~\ref{fig:16A} & -0.57 &  0.60 & 0.103 & -0.45 &  0.46 & 0.080 & -0.15 &  0.08 & 0.017  &  0.69 \\  %  16A
%Fig.~\ref{fig:16B} & -0.27 &  0.40 & 0.098 & -0.39 &  0.32 & 0.088 & -0.07 &  0.09 & 0.016  &  1.43 \\  %  16B
    \end{tabular}
    \caption{Summary of the largest and smallest deflection angles, and the corresponding rms values, as well as the largest and smallest values and the rms of $\Delta\kappa$ of the ShaDes presented in the Figures. The last column gives the median percent difference (MPD) between the free-form {\tt Grale} model of MACS J0416 presented in \cite{perera2024b} and {\tt Grale} with ShaDes added. MPD is evaluated over the rectangular area shown with gray dashed lines in the right panel of Figure~\ref{fig:07F}.}
    \label{tab:summary}
\end{table*}

\subsection{Effect of removing tight image groupings}\label{sec:tight}

\begin{figure*}
    %\vspace{-4cm}
    \centering
    \includegraphics[trim={0.7cm 5cm 1cm 5cm},clip,width=0.32\linewidth]{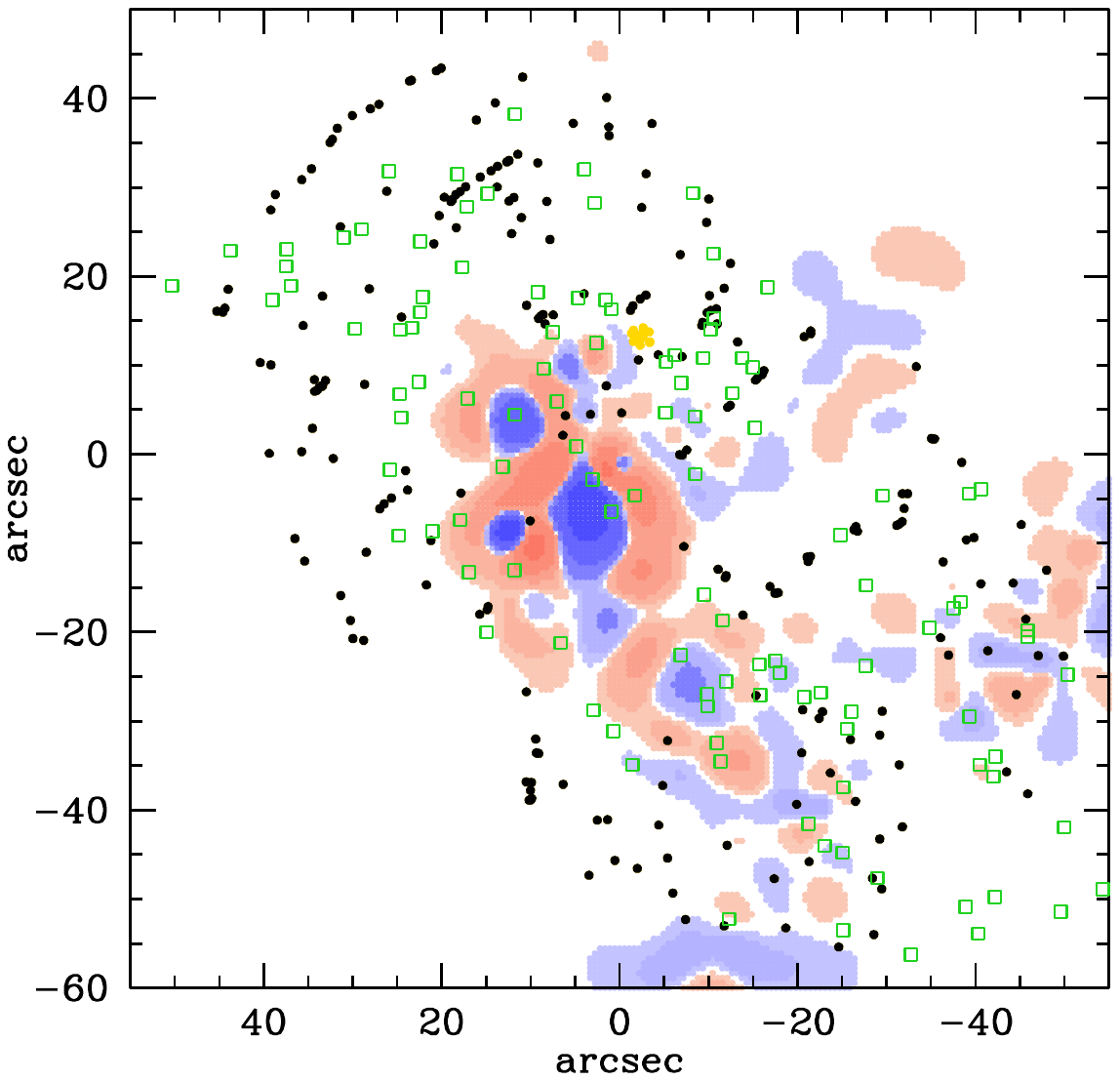}  %% 
    \includegraphics[trim={0.7cm 5cm 1cm 5cm},clip,width=0.32\linewidth]{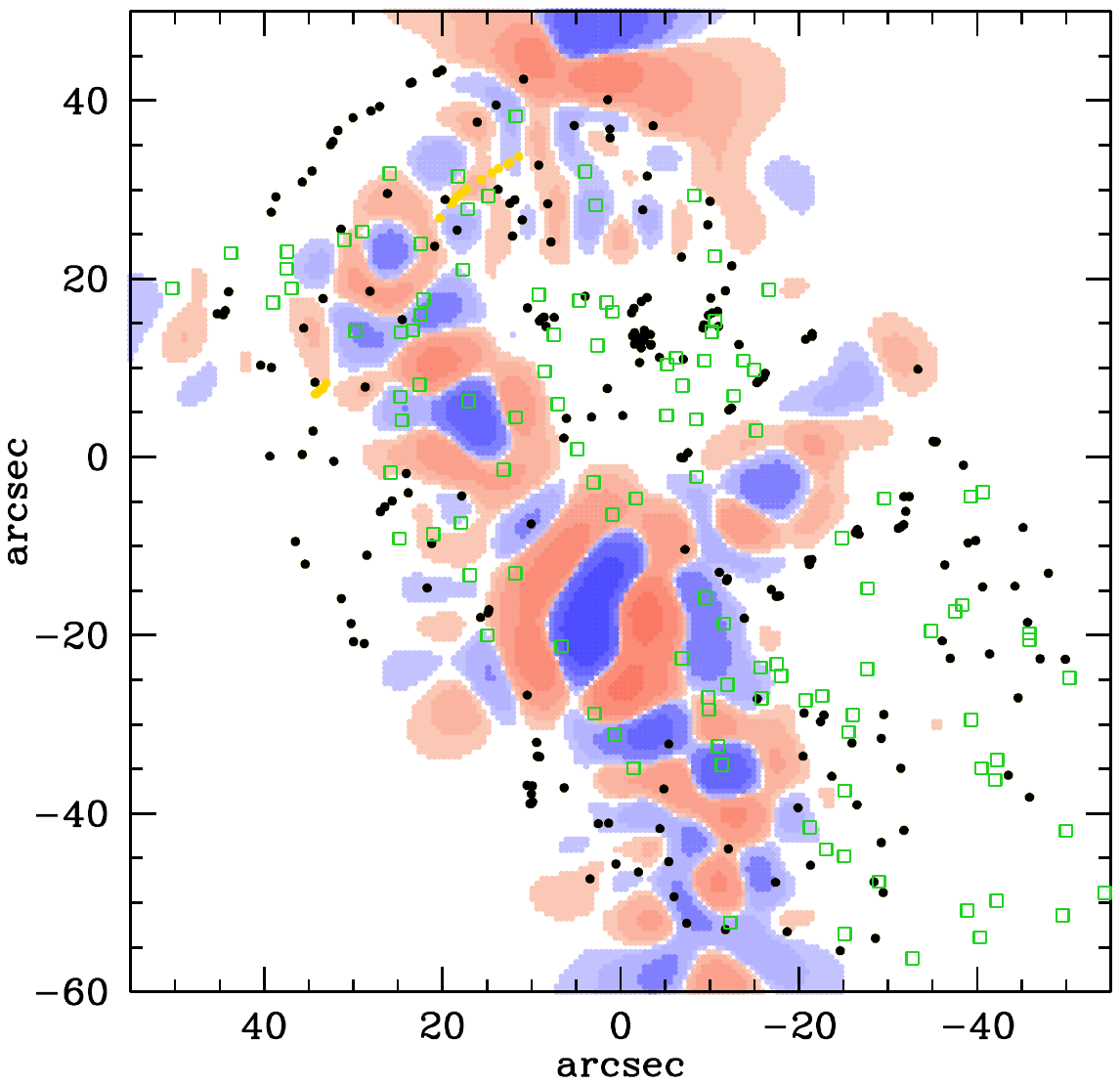}
    \includegraphics[trim={0.7cm 5cm 1cm 5cm},clip,width=0.32\linewidth]{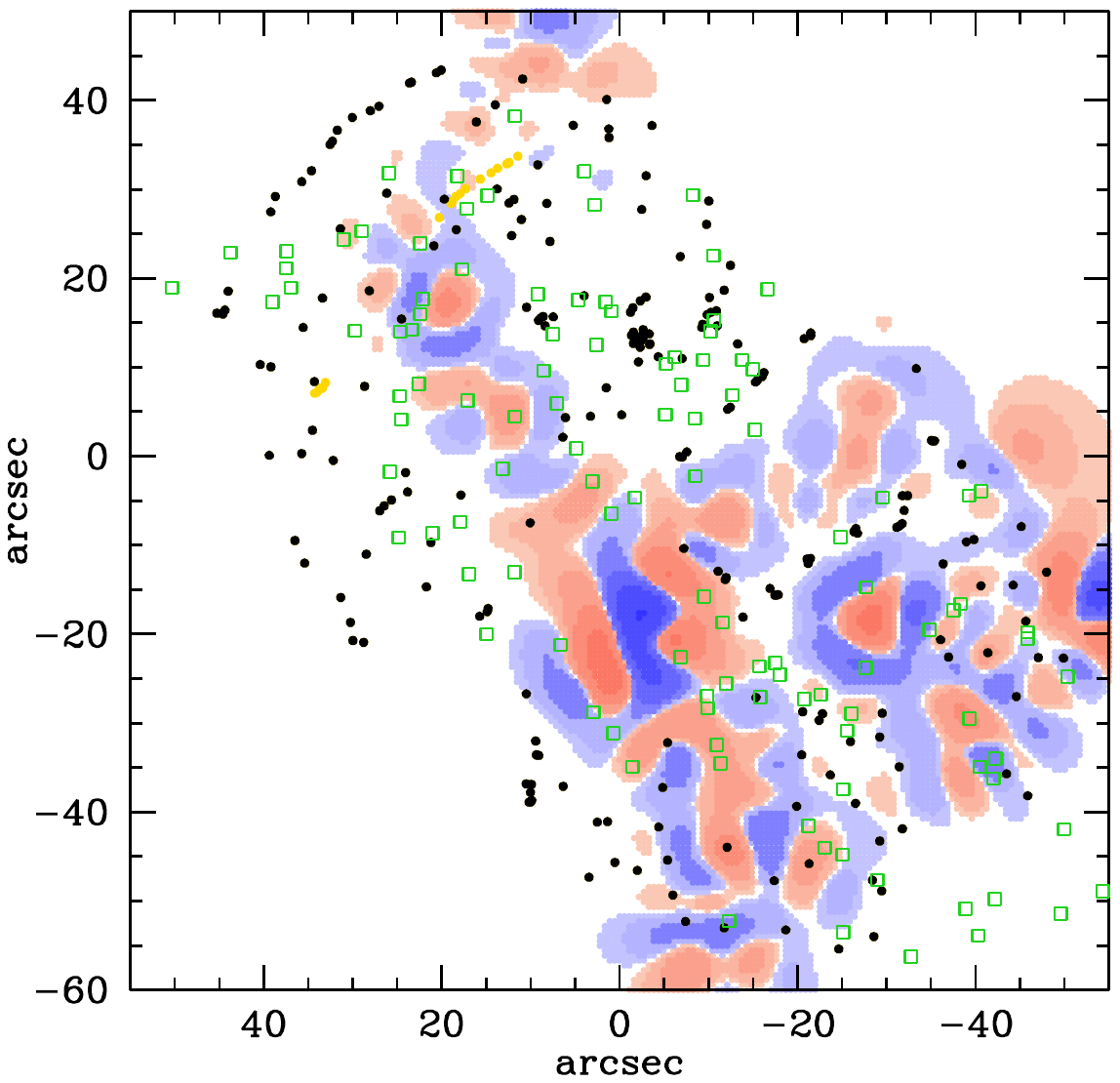}
    %%\vspace{-1cm}
    \caption{Similar to $\Delta\kappa$ maps shown in the right panels of Figures~\ref{fig:07F}-\ref{fig:07D},  but some sets of images were removed from the analysis. {\it Left:} Twelve images of the Warhol arc were removed, around $(-3'',12'')$ (shown in yellow). {\it Middle:} Eighteen images of the arc around $(15'',30'')$ and around $(35'',7'')$ were removed (shown in yellow). {\it Right:} Similar to the middle panel, but a different realization. 
    %See Table~\ref{tab:summary} for values characterizing deflection angles and $\Delta\kappa$.
    }
    \label{fig:1106}
\end{figure*}

The above finding that dense groupings of images likely lead to suppression of shape degeneracies suggests an experiment. We carry out ShaDes analysis after removing 12 images of the Warhol arc, leaving us with 225 images. The left panel of Figure~\ref{fig:1106} shows one realization, with the removed images highlighted in yellow. It is apparent that the region near these missing images is now more populated with ShaDes, compared to when these images were included. Even so, ShaDes introduce only small density perturbations at the location of "removed" Warhol. We attribute this to the presence of other, smaller groupings of images in the same $10''-15''$ region, which were not removed in this experiment.

In the next two panels of the same Figure, 18 images of the arc around $(15'',30'')$ and $(35'',7'')$ were removed (images shown in yellow), and Warhol images added back. Now ShaDes cover the region with the yellow points, but avoid the Warhol region. The regions in the lens plane with 18 removed images are better populated by ShaDes density perturbations than in the above experiment with Warhol. This is probably because the extended region around these 18 images hosts only isolated images, not tight groupings.
Table~\ref{tab:summary} summarizes the ranges of deflection angles and $\Delta\kappa$ for these ShaDes. 

These experiments support our claim that dense groupings of images strongly suppress shape degeneracies. 

\subsection{Mass clump M2}\label{sec:m2}

Free-form reconstruction of \cite{perera2024b} recovered a mass clump, dubbed M2, not affiliated with any light. Its mass was measured to be $5.7\pm0.2\times 10^{11}\,M_\odot$, within $r=8\,$kpc, or $1.5''$ from its peak. \cite{Limousin2025} performed several parametric and free-form reconstructions of MACS J0416. While all reconstructions agree in general, the details differ. Free-form {\tt Grale} detects M2 as an amorphous mass clump, while some models show a mass extension in that direction, and some others do not register anything unusual near its location. 

For example, the surface mass density around M2 is in reasonable agreement between {\tt Grale} and {\tt Lenstool} (they agree within $5-6\%$), better than over the rest of the cluster (see the bottom panel of their Figure~3)\footnote{Interestingly, the largest disagreement in density in that region is between two {\tt Lenstool} models: \cite{Limousin2025} (Appendix A), and \cite{rihtarsic2025}; see the bottom panel of \cite{Limousin2025}'s Figure~3.}. However, the two models disagree on the density compactness, with {\tt Grale} favoring a distinct mass substructure, while {\tt Lenstool} preferring a more diffuse mass distribution.

\cite{Limousin2025} performed a further test of M2. A mass component was added to {\tt Lenstool} at the location of {\tt Grale}'s M2 with initial parameters similar to those of {\tt Grale}'s M2, and the {\tt Lenstool} model was optimised as usual (see their Section~4.2). The result is that {\tt Lenstool} prefers to remove this additional mass component, i.e., its amplitude ends up being zero, implying that {\tt Lenstool} model is arguing against a compact mass subclump at the location of M2.

\begin{figure*}
    %\vspace{-4cm}
    \centering
    \includegraphics[trim={0.7cm 5cm 1cm 5cm},clip,width=0.32\linewidth]{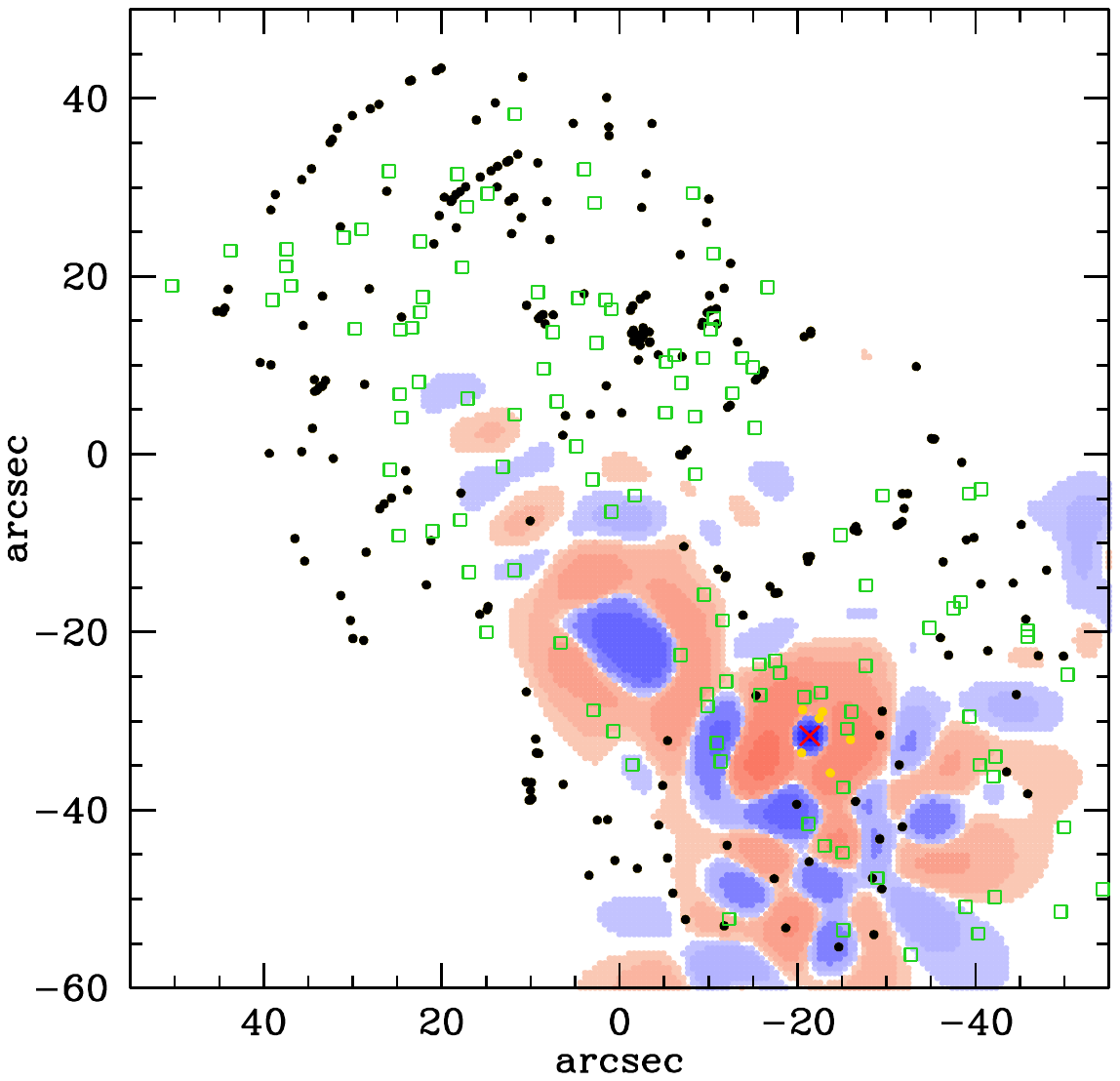}  
    \includegraphics[trim={0.7cm 5cm 1cm 5cm},clip,width=0.32\linewidth]{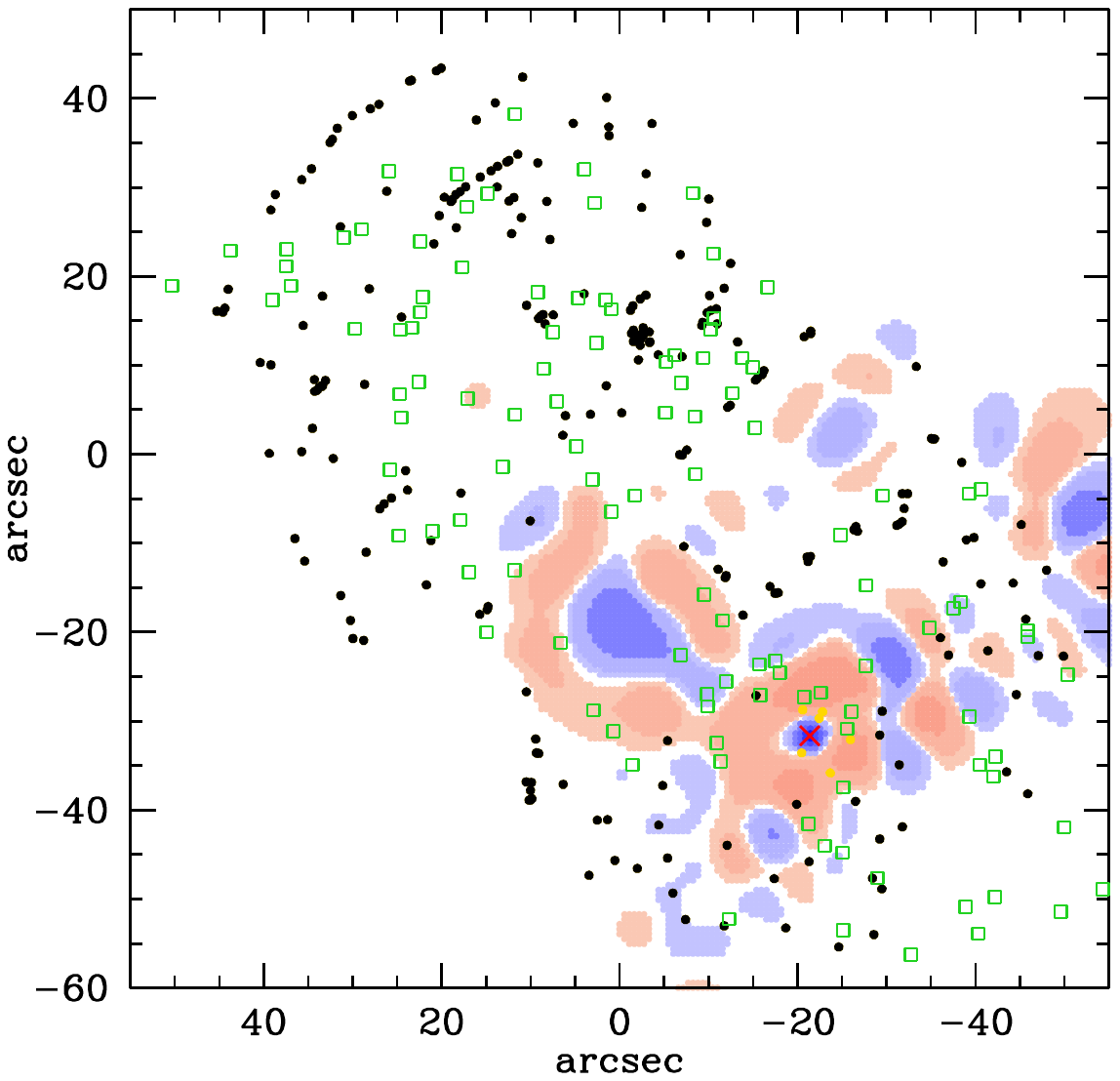}      
    \includegraphics[trim={0.7cm 5cm 1cm 5cm},clip,width=0.32\linewidth]{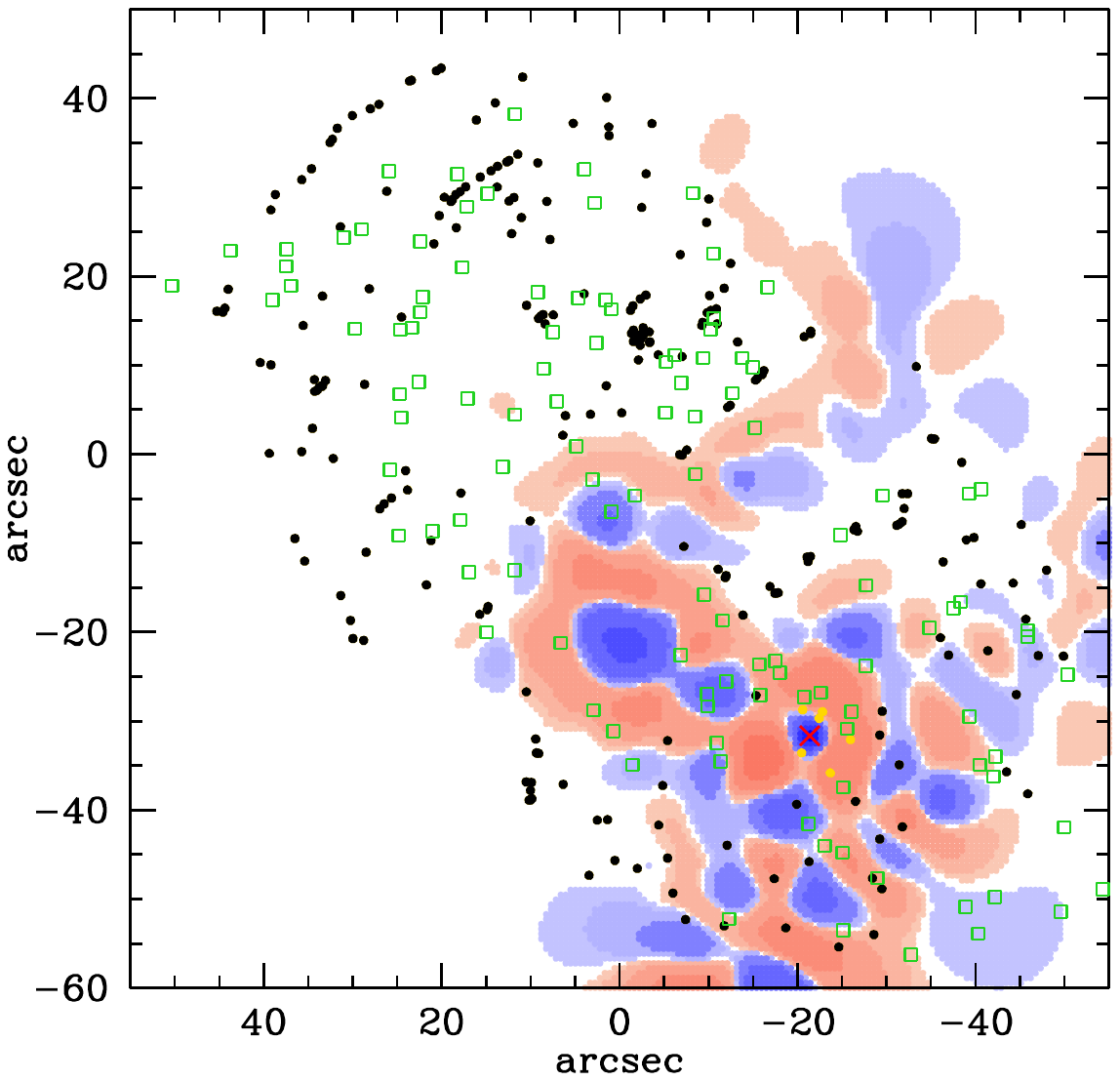}  
    \caption{Similar to $\Delta\kappa$ maps shown in the right panel of Figure~\ref{fig:07F}, but here we generate ShaDes that when added to a mass model would "remove" M2, whose location is marked with a red cross. Blue regions have $\Delta\kappa<0$, so when added to a {\tt Grale} model, M2 will nearly disappear, mimicking {\tt Lenstool} reconstructions of MACS J0416.
    %See Table~\ref{tab:summary} for values characterizing deflection angles and $\Delta\kappa$.
    }
    \label{fig:14}
\end{figure*}

In this paper we ask if the differences between recovered mass distributions of different models around the location of M2 can be accounted for by ShaDes. {\tt Grale} is taken to be model $\mathcal{M}(\bm{\theta})$, which contains M2 mass clump. Therefore to use ShaDes to make a different model, which does not have M2 as a localized clump, $\mathcal{P}(\bm{\theta})$ must have a crater with $\Delta\kappa<0$, and total (negative) mass and extent similar to that of {\tt Grale}'s M2 at the location of M2, to cancel it out. To make that happen we place a negative density fixed alphapot at M2, and do not allow any basis functions to be centered closer than $1.1''$ from the center of M2. This distance and properties of the fixed alphapot result in a crater in $\mathcal{P}(\bm{\theta})$ that compensates for M2. Figure~\ref{fig:14} shows 3 different realizations of such ShaDes, with M2 marked by a red cross near $(-22'',-30'')$.

Comparing the three degenerate models $\mathcal{S}_i({\bm{\theta}})=\mathcal{M}({\bm{\theta}})+\mathcal{P}_i({\bm{\theta}})$ to the one named L25 from \cite{Limousin2025}, first note that L25 reconstructs the positions of 3 images located within $3.06''$ from M2, to within $0.13'' - 0.78''$ from their observed positions. The $\mathcal{M}(\bm{\theta})$ model from {\tt Grale} reconstructs these images to $0.03'' - 0.14''$, which is unaffected by the addition of $\mathcal{P}_i(\bm{\theta})$ if the images are included in the procedure to build $\mathcal{P}_i(\bm{\theta})$. To allow the offsets in the $\mathcal{S}_i(\bm{\theta})$ models to increase to the size in the L25 model, we now leave out the images within $5''$ from M2; the 6 images within this radius include the 3 images mentioned before. ShaDes are then constructed using the remaining 231 images.

At $1.5''$ from the M2 center, the typical deflection angle in our ShaDes $\mathcal{P}_i(\bm{\theta})$ maps, is $0.6''$. (Since there are no images in this region, deflection angles due to $\mathcal{P}_i(\bm{\theta})$ are non-zero.) This is comparable to the displacement of image K54.2, which is $0.65''-0.78''$, as predicted by L25. This means that our constructed ShaDes, when added to a {\tt Grale} model can ``remove" M2 at the expense of increasing the distance between observed and reconstructed images by $\sim 0.6''$ in the vicinity of M2. Thus our constructed $\mathcal{S}_i(\bm{\theta})$'s are consistent with L25 in the vicinity of M2.

We conclude that the ``Schrodinger Cat" behavior of M2 can be accounted for by lensing shape degeneracies. This conclusion does not confirm or negate the physical reality of M2. More lensed images in the region of purported M2 would need to be found and modeled before M2's reality is addressed. \new{The same caveat applies to other dark clump candidates reported in the literature using free-form inversions \citep{ghosh2023,perera2024a}.}

\subsection{Effect on the determination of $H_0$}\label{sec:h0}

\begin{figure}
    %\vspace{-4cm}
    \centering
    \includegraphics[trim={0.7cm 5cm 1cm 5cm},clip,width=0.95\linewidth]{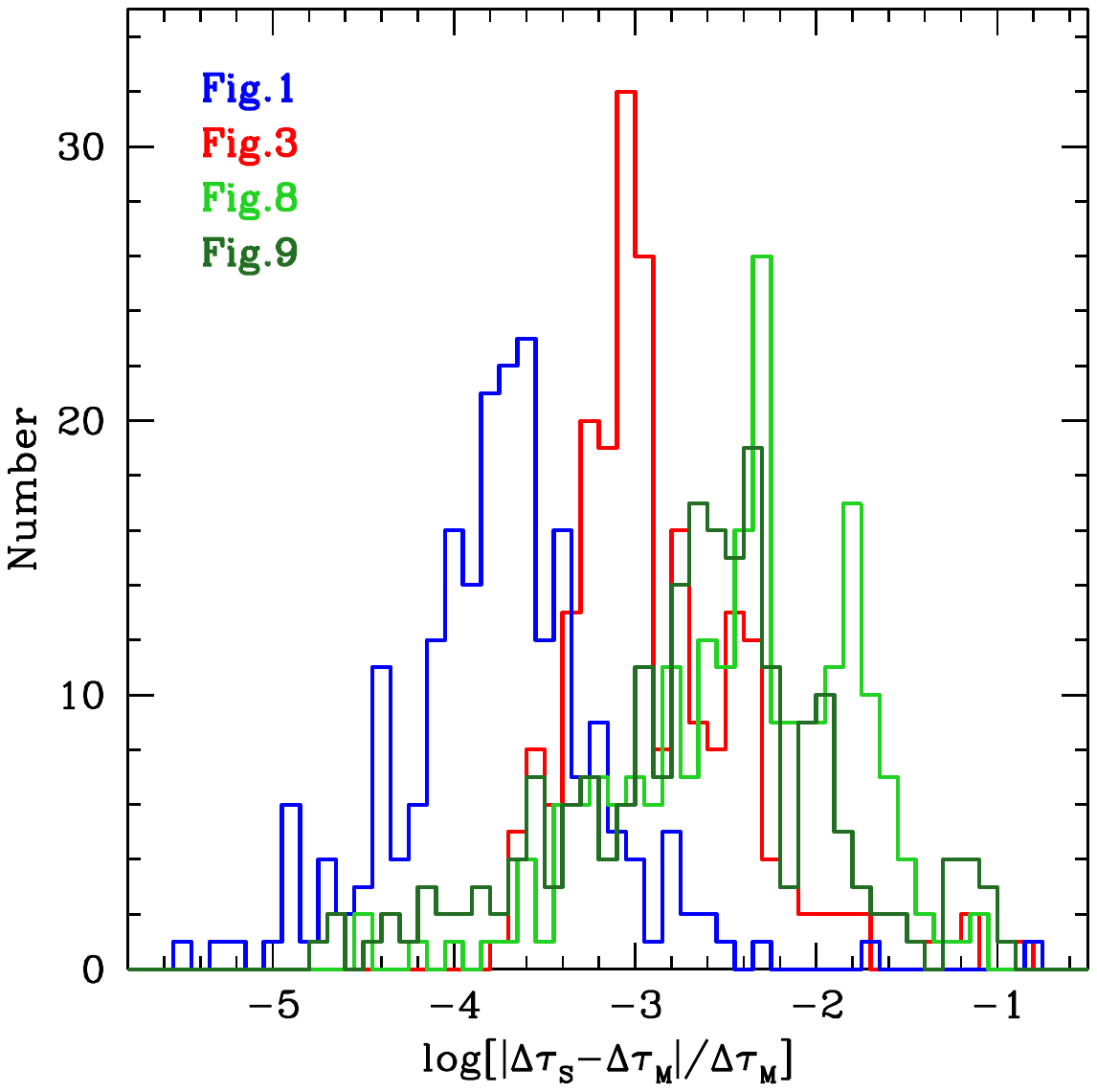}  
    \caption{Histogram of the logs of the fractional difference in time delays when shape degeneracies are added to an existing model, i.e., $\mathcal{S}_i(\bm{\theta})=\mathcal{M}(\bm{\theta})+\mathcal{P}_i(\bm{\theta})$. The mass distribution and the reconstructed sources of the free-form {\tt Grale} are used as $\mathcal{M}(\bm{\theta})$. The four ShaDes $\mathcal{P}_i(\bm{\theta})$ we use are presented in Figures as indicated in the legend. }
    \label{fig:H0}
\end{figure}

Since the detection of Supernova Refsdal \citep{kelly2015}, clusters that host multiple images of supernovae or quasars with measured time delays have been used to estimate the Hubble constant, $H_0$, using
\begin{equation}
\begin{split}
    \tau(\bm\theta)=
    &\frac{(1+z_l)}{c}\frac{D_{ol}D_{os}}{D_{ls}}\times\\
    &\Big[\frac{1}{2}(\bm\theta-\bm\beta)^2-\frac{1}{\pi}\int d\bm\theta'^2\kappa(\bm\theta')\ln\frac{|\bm\theta-\bm\theta'|}{\theta_0} \Big],\label{eq:H0}
\end{split}
\end{equation}
\citep[e.g.,][]{napier2023}, thereby realizing the vision of Sjur Refsdal \citep{refsdal1964}, 50 years later.
Here we ask if ShaDes have an effect on this determination of $H_0$ in clusters. Since MACS J0416 has no known supernova or quasars, we pretended that all images of the same source had measured time delays. We then calculated the fractional difference in arrival time of all two-image pairs from same sources.  Using all 88 sources this gave us 217 pairs of images. 

We took the two ShaDes shown in Figures~\ref{fig:07F} and \ref{fig:07D}, $\mathcal{P}_i(\bm{\theta})$, added them to the {\tt Grale} \citep{perera2024b} mass model ($\mathcal{M}(\bm{\theta})$), thereby creating two versions of $\mathcal{S}_i(\bm{\theta})$. 
Figure~\ref{fig:H0} shows the histogram of the log of the absolute fractional time delay difference for these two $\mathcal{S}_i(\bm{\theta})$ models in blue and red, respectively. Looking at eq.~\ref{eq:H0} we see that these fractional differences translate directly into the error in estimated $H_0$. The peaks of these two distributions are at $\lesssim 1\%$, and for a very small fraction of cases that error can rise up to almost $10\%$ (red histogram). A few percent error is comparable to, but typically smaller than the uncertainty on the measured time delays. This level of uncertainty will not dominate the error budget of $H_0$ estimation. \new{Note that we are applying ShaDes to a cluster with over 200 images; the percent differences quoted here may not apply to clusters with much fewer images.}

We will discuss the other two histograms shown in this Figure (light and dark green) in Section~\ref{sec:shades+}, where we account for the fact that lens models' do not reproduce the positions of the observed multiple images exactly.

%\section{Beyond S\lowercase{ha}D\lowercase{es}}\label{sec:beyond}

\subsection{Effect of offsetting sources}\label{sec:offset}

\begin{figure*}
    %\vspace{-4cm}
    \centering
    \includegraphics[trim={0.7cm 5cm 1cm 5cm},clip,width=0.32\linewidth]{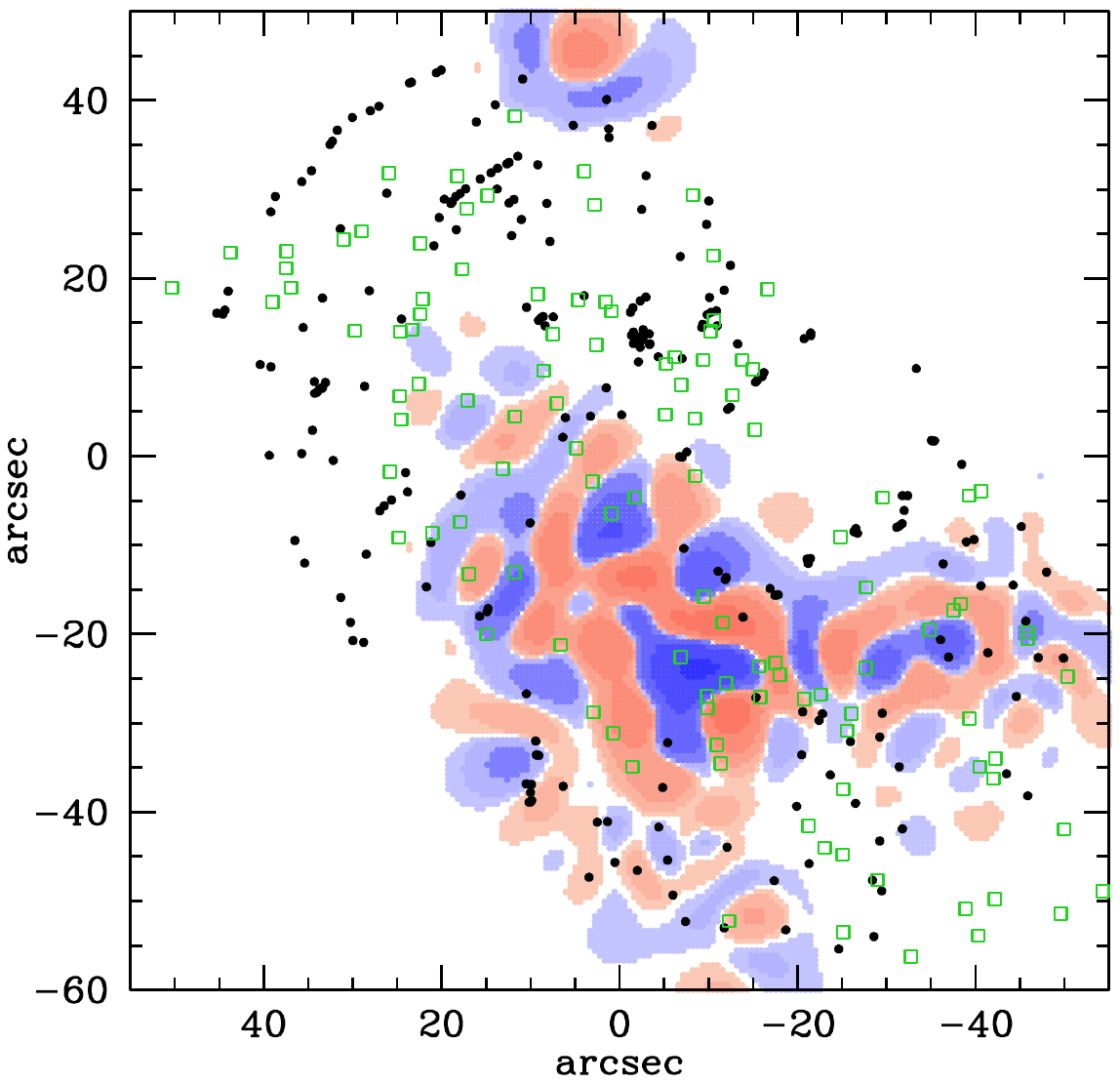}  %% soff=0.2
    %%\vspace{-1cm}
    \includegraphics[trim={0.7cm 5cm 1cm 5cm},clip,width=0.32\linewidth]{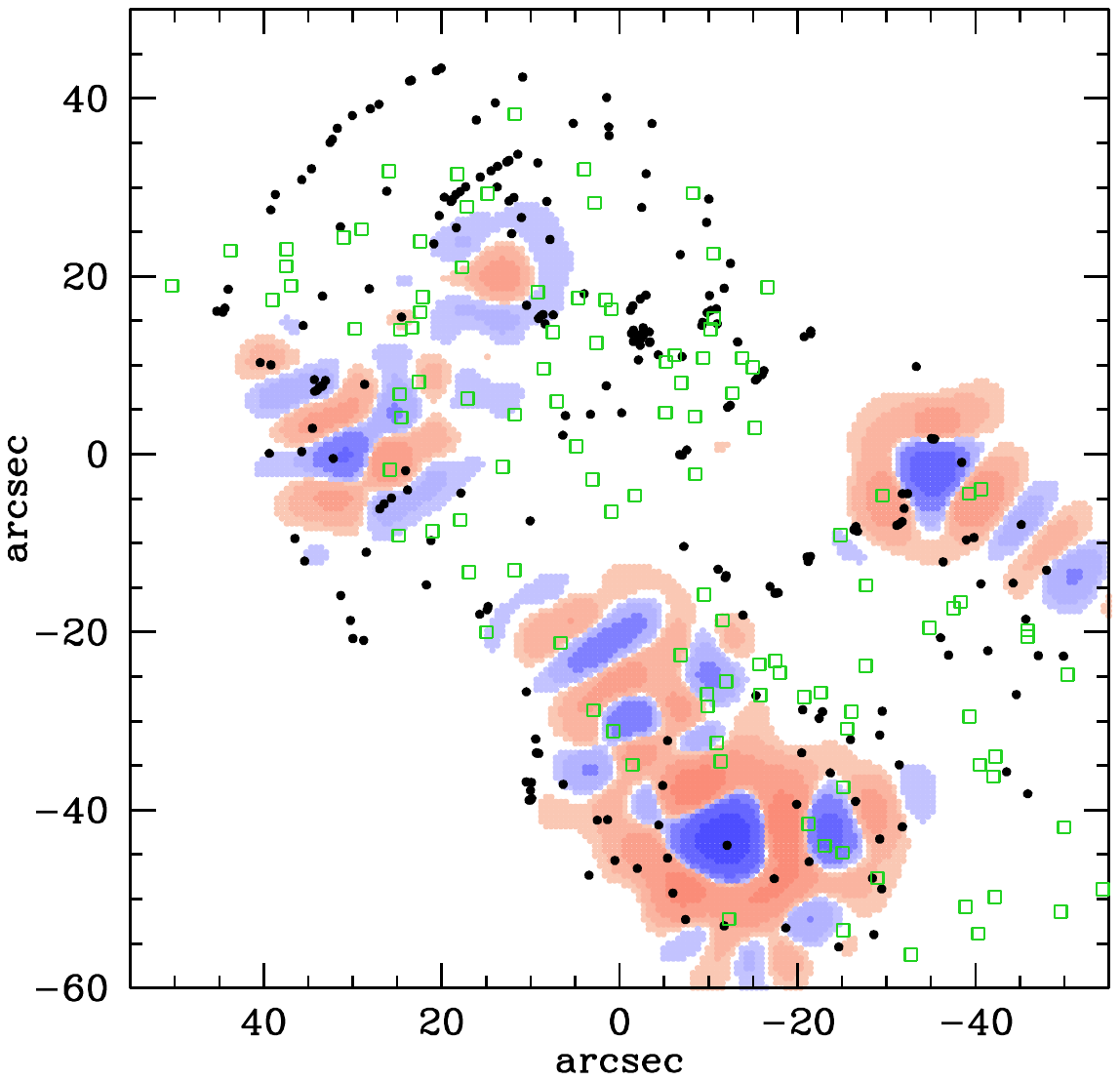}  %% soff=0.4
    \includegraphics[trim={0.7cm 5cm 1cm 5cm},clip,width=0.32\linewidth]{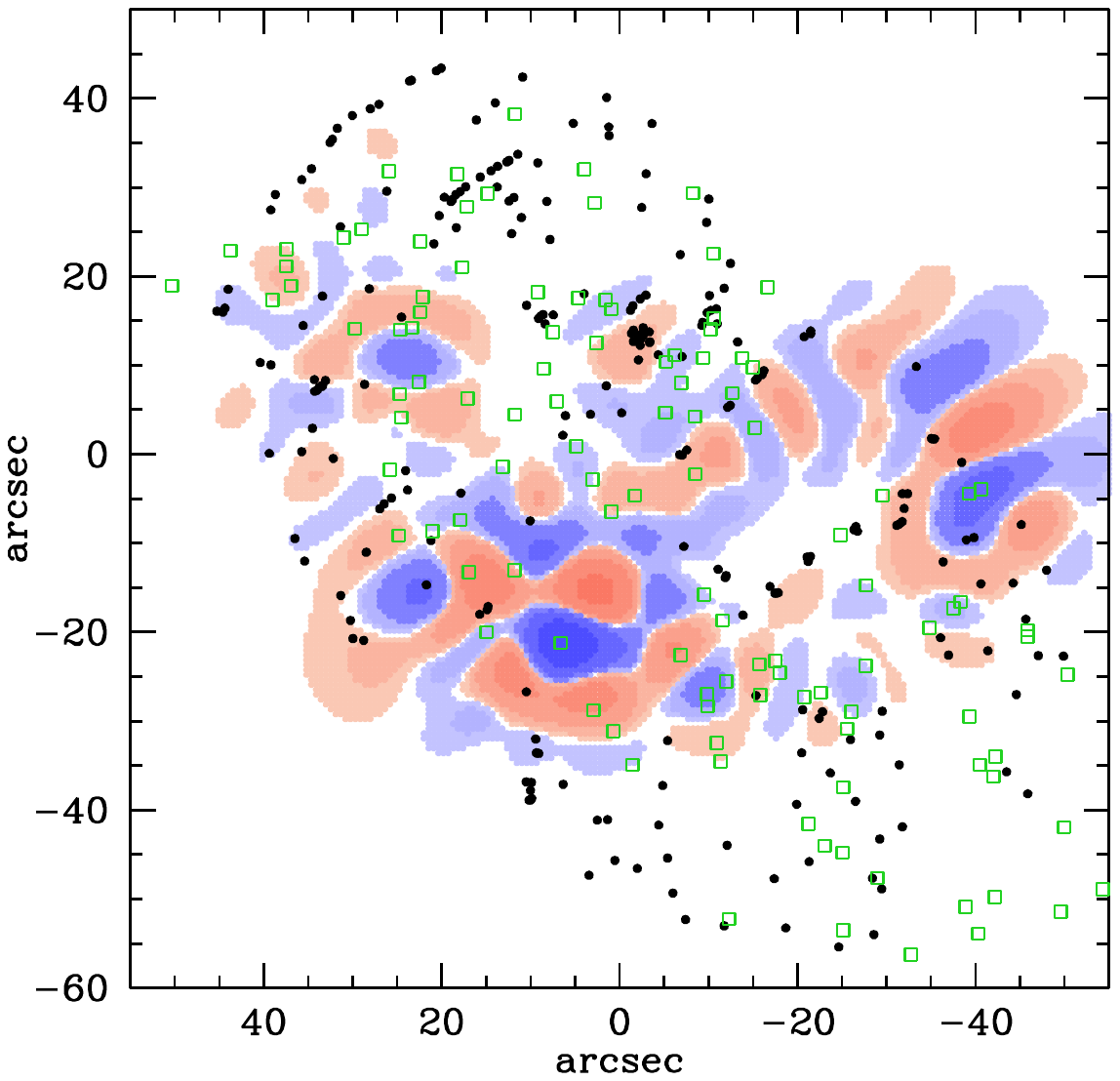}  %% soff=0.4
    \caption{Similar to $\Delta\kappa$ maps shown in the right panel of Figure~\ref{fig:07F}, but sources were offset by a random amount between $0''$ and $22''$ (left panel), and $0''$ and $44''$ (middle and right panels). 
    %See Table~\ref{tab:summary} for values characterizing deflection angles and $\Delta\kappa$.
    }
    \label{fig:08}
\end{figure*}
According to our definition of ShaDes, the deflection angles at the image positions are preserved, and since the image positions also have to be preserved, that means the source positions are unchanged from their original ones recovered by $\mathcal{M}(\bm{\theta})$.

However, the linear algebra mechanism we use to generate ShaDes also allows us to study a different degenerate transformation. We can allow sources to be displaced from their model predicted positions by some amount, so that each source and its images are described by $\bm{\theta}=[\bm{\beta}+\bm{\delta}(z_s)]+[\bm{\alpha}(\bm{\theta})-\bm{\delta}(z_s)]$. We choose $\bm\delta(z_s)$ to be constant for all sources at a given redshift, but it can differ between source planes. The same deflection angle applied to all images of sources at the same redshift is equivalent to having a wedge-like potential, $\Psi= a(z_s)x+b(z_s)y$, added just in front of these sources. Here, $a(z_s)$ and $b(z_s)$ are constants for any given source redshift, and $x$ and $y$ are coordinates in the source plane(s). The deflection angles will be constant across the field, and the added surface density is zero everywhere. Such an idealized construct is analogous in some ways to external shear, $\Psi=\frac{1}{2}\gamma (x^2-y^2)$, and can also imitate distant masses. These mass concentrations are due to the large scale structure in the Universe, and exist at all redshifts. There is ample evidence for such structures in the redshift surveys in the directions of clusters \citep{bayliss2014,lagattuta2022}, and there may be indirect evidence in lens reconstructions \citep{williams2018}. Their effect can be represented by wedge-like lensing potentials that have different orientations and amplitudes encoded in $a(z_s)$ and $b(z_s)$, at a range of redshifts, $z_s$.

Therefore we now relax the assumption that source positions are the same as in the original unperturbed model $\mathcal{M}$, and randomly displace images belonging to sources at the same redshift by ${\bm\delta}(z_s)$. All images that share exactly the same redshift are considered to belong to the same source plane. There are 56 sources planes in MACS J0416. %We do this so that all images originating from the same redshift would have the same source plane offset. 
In the first panel of Fig.~\ref{fig:08} we pick source displacements randomly between $0''$ and $22''$, while the middle and right panels allow displacements as large as $44''$ in the source plane. These displacements are probably unrealistically larger, but we wanted to explore the extreme version of this scenario. We use all 237 images in all these models.

The lens plane regions that were avoided by ShaDes in Fig.~\ref{fig:07F}-\ref{fig:07D} now do have non-zero $\Delta\kappa$. However, in reality the source displacements are likely smaller than assumed here,  therefore tight groupings of images remain an effective guard against these degeneracies as well.

\section{Effect of non-zero lens plane rms; ShaDes+}\label{sec:shades+}

By our definition, shape degeneracies, or ShaDes preserve the deflection angles at the locations of known observed images. In that sense they are exact degeneracies. However, no two lens reconstructions recover exactly the same image positions, and most reconstructions do not reproduce observed images to angular resolution of the observations. 
To account for this reality we define ShaDes+ as {\it approximate} shape degeneracies, which do not preserve image positions exactly, but allow for some offset, comparable to that seen in reconstructions. We call these ShaDes+, because they afford more freedom than ShaDes.

For example, the differences between the models presented by \cite{bergamini2021} and \cite{Bergamini2023}, are shown in Figure~6 (left panel) of the latter paper. The $\Delta\kappa$ contours cannot be described by simple parametric forms, typical percent difference between the two $\kappa$ maps is $\lesssim 4\%$, and the images are not at exactly the same positions in the two reconstructions, implying that these two maps are related by ShaDes+. %LPrms is $0.4''$ and $0.43''$. 

Figure~4 of \cite{rihtarsic2025} displays the fractional difference in $\kappa$ between their model and that of \cite{Bergamini2023}. The amorphous nature of the positive and negative difference regions across the cluster and the fact these two models do not reproduce exactly the same image positions means that this is another example of ShaDes+.\footnote{The number of images used in \cite{bergamini2021}, \cite{Bergamini2023} and \cite{rihtarsic2025}  models is not the same, but because ShaDes+ are not tied to the number of images used, these are examples of ShaDes+.} 

Compared to ShaDes, ShaDes+ are more representative of the difference between lens models found in the literature. As with all other lensing degeneracies, it is the approximate ones that are more relevant that the exact ones.

To construct ShaDes+ we still use the linear algebra formalism, but the locations where deflection angles are preserved are now not the observed image positions, but random positions in the cluster. This allows for deflection angles at the observed images to differ from zero, reflecting the actual situation with models. The main role of the linear algebra construction is to keep $|\Delta\kappa|$ in check across the cluster, not letting it get too large. 

Since we are now not tied to the observed number of images, we can choose to have deflection angles be zero at fewer locations, thereby increasing the linear scale of resulting $\Delta\kappa$ regions.

\begin{figure*}
    %\vspace{-4cm}
    \centering
    \includegraphics[trim={0.7cm 5cm 1cm 5cm},clip,width=0.32\linewidth]{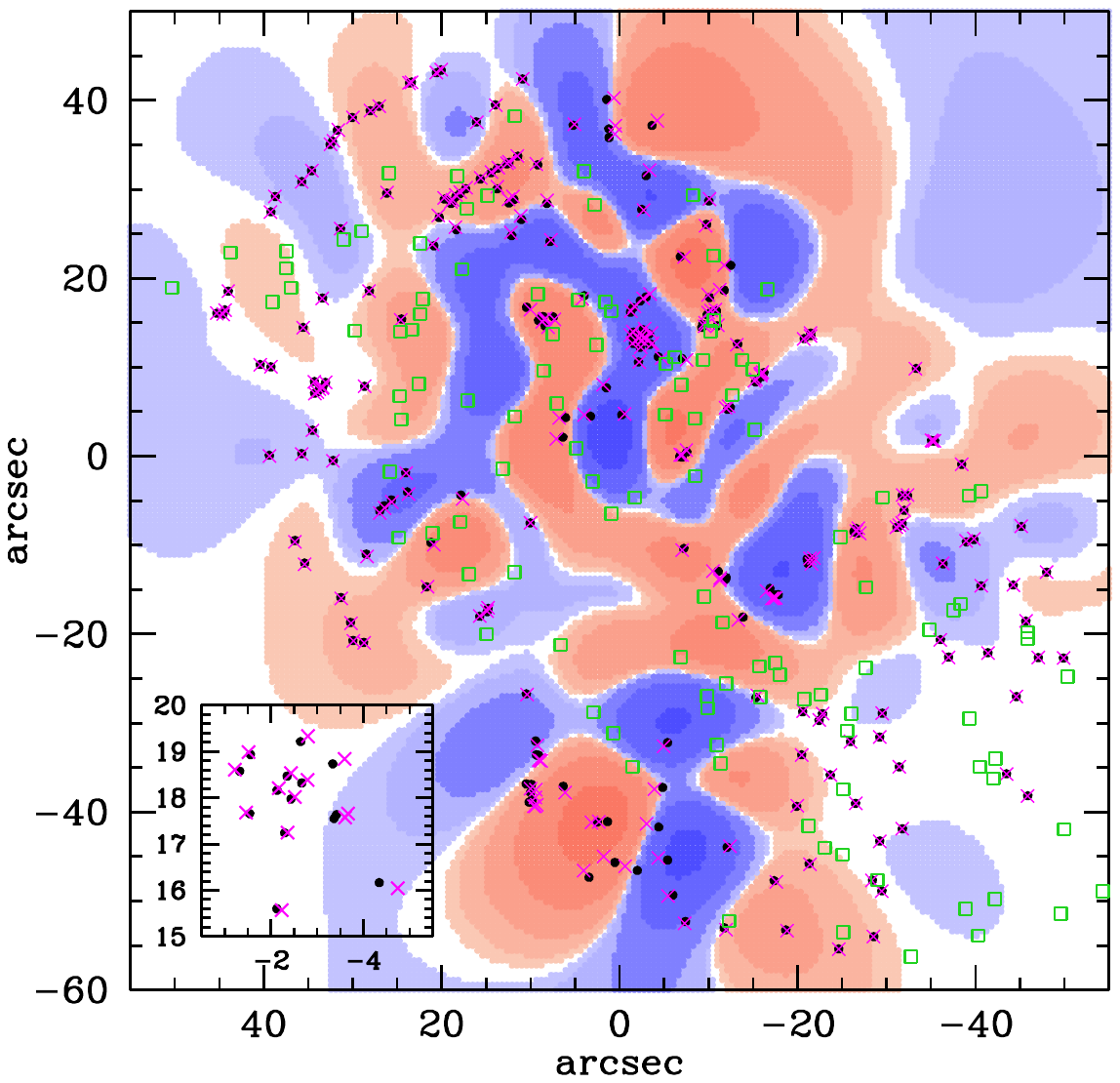} 
    %%\vspace{-1cm}
    \includegraphics[trim={0.7cm 5cm 1cm 5cm},clip,width=0.32\linewidth]{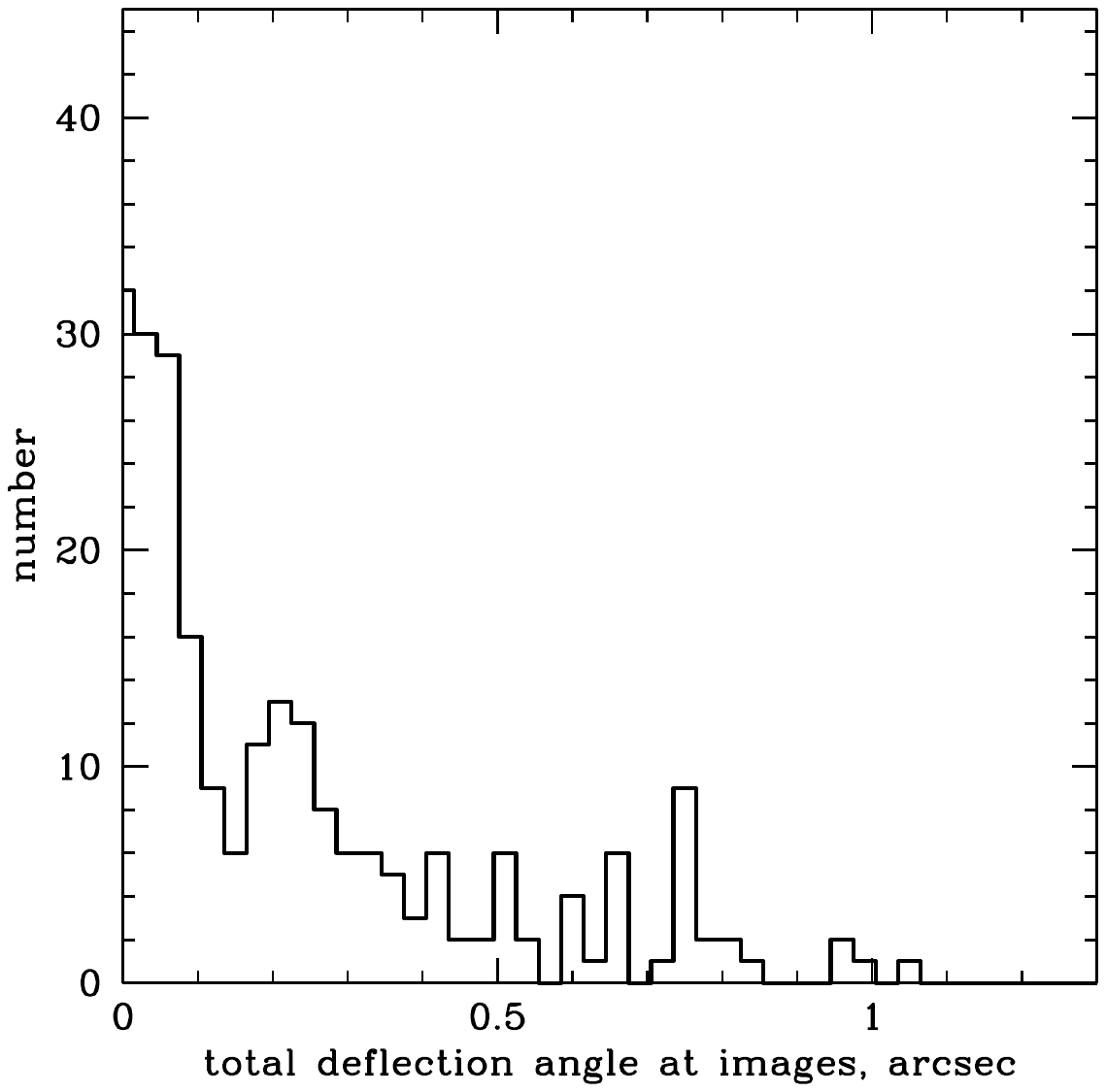}  
    \includegraphics[trim={0.7cm 5cm 1cm 5cm},clip,width=0.32\linewidth]{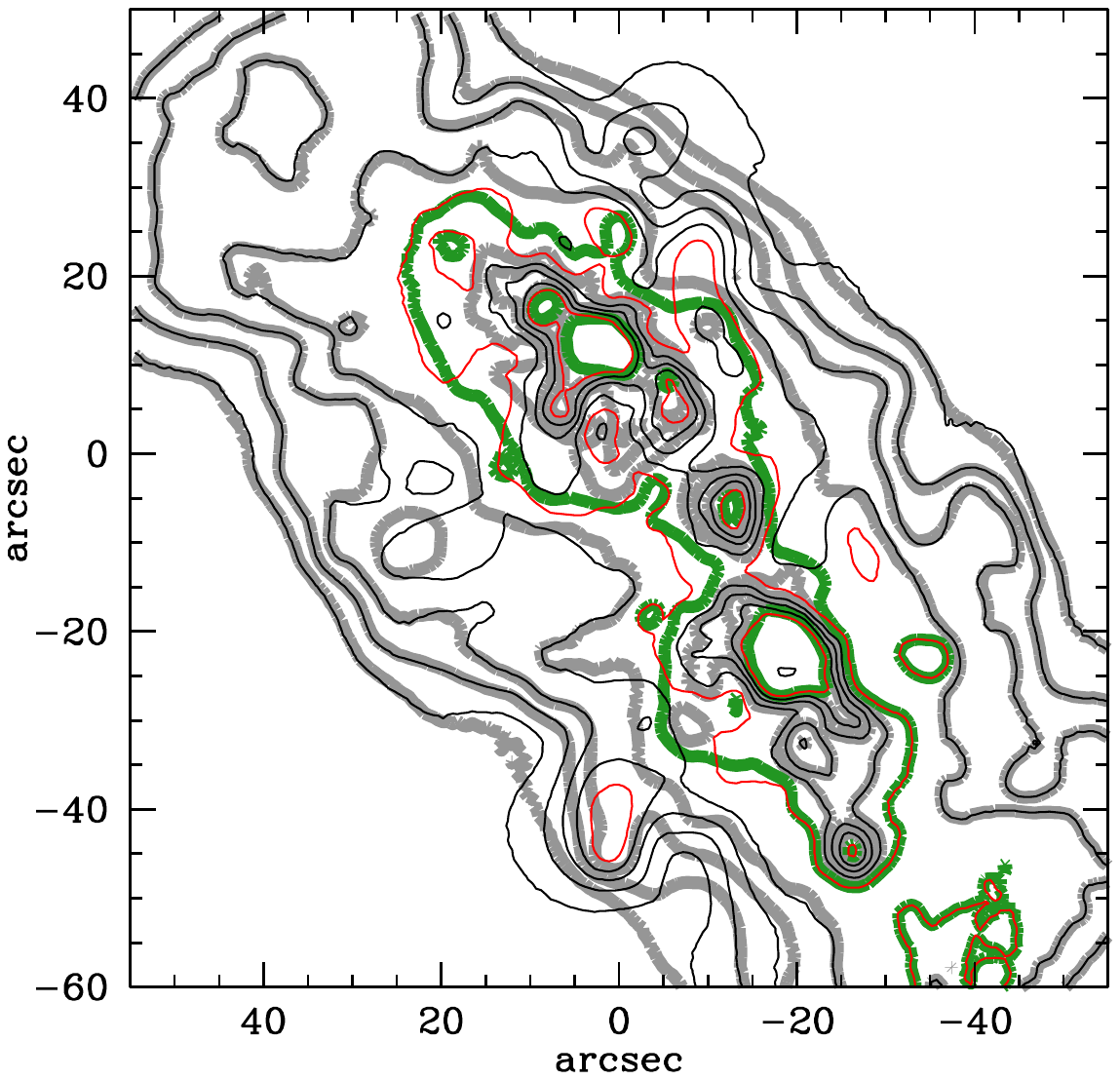} 
    \caption{An example of ShaDes+, where the observed images are the black dots (left panel). When the deflection angles at their locations are applied the resulting positions are marked with magenta crosses. The inset shows a zoom-in of the Warhol region, where the difference between the two is more visible. The middle panel shows a histogram of offsets between observed images and those with ShaDes+ deflections applied. The maximum offset is $\sim0.65''$, and the average is smaller than typical lens plane rms of reconstructions. The right panel shows {\tt Grale}'s $\kappa$ map of MACS J0416 as thick gray ($\kappa=0.1, 0.2, 0.3, 0.4, 0.6, 0.7, 0.8, 1.3$) and green lines ($\kappa=0.5$ and $0.9$). When ShaDes+ are applied the new contours, at the same $\kappa$ levels as shown as thin black and red lines. The contours do not change much, and there are no regions with $\kappa<0$.
    }
    \label{fig:17A}
\end{figure*}

\begin{figure*}
    %\vspace{-4cm}
    \centering
    \includegraphics[trim={0.7cm 5cm 1cm 5cm},clip,width=0.32\linewidth]{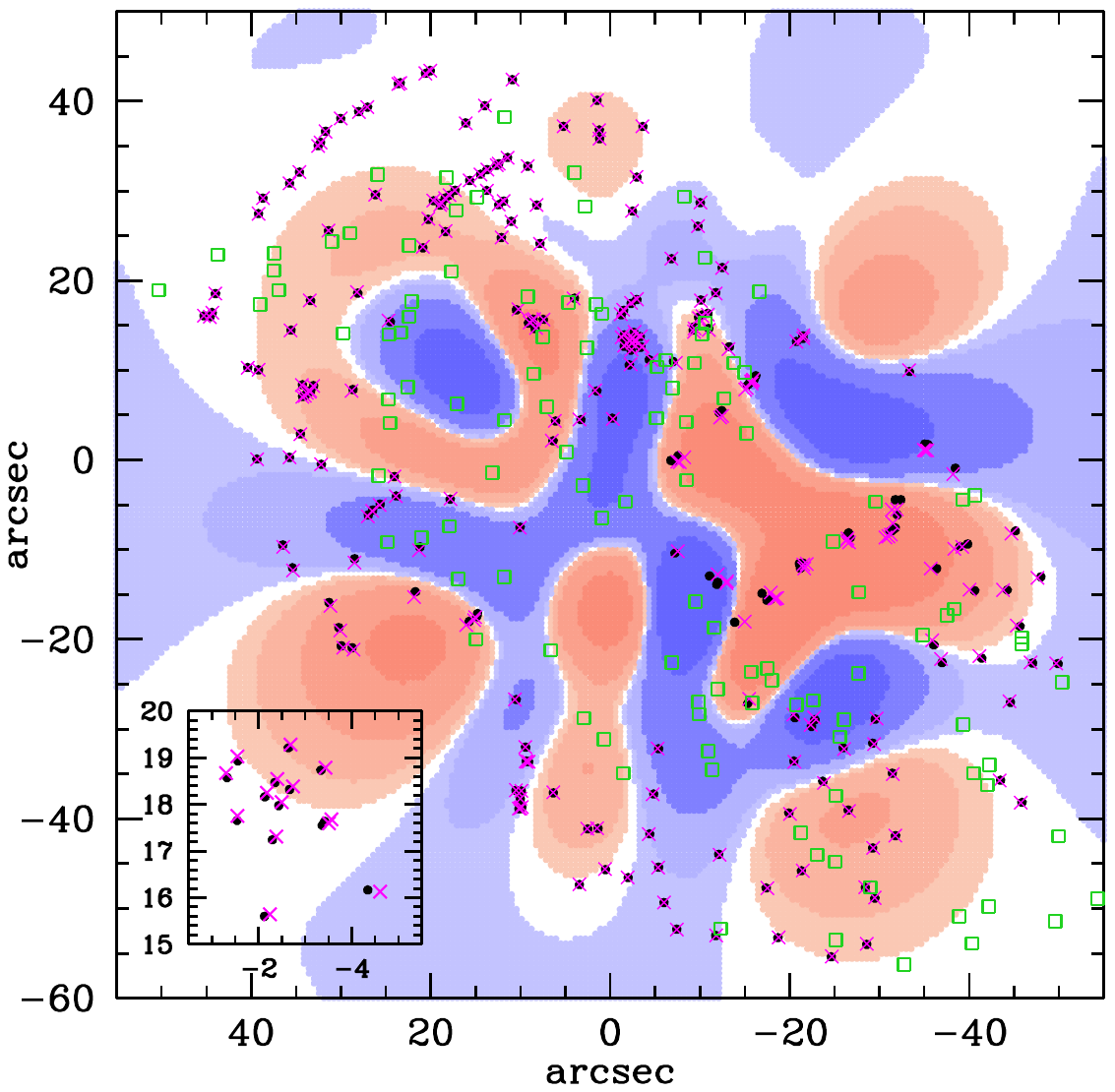} 
    %%\vspace{-1cm}
    \includegraphics[trim={0.7cm 5cm 1cm 5cm},clip,width=0.32\linewidth]{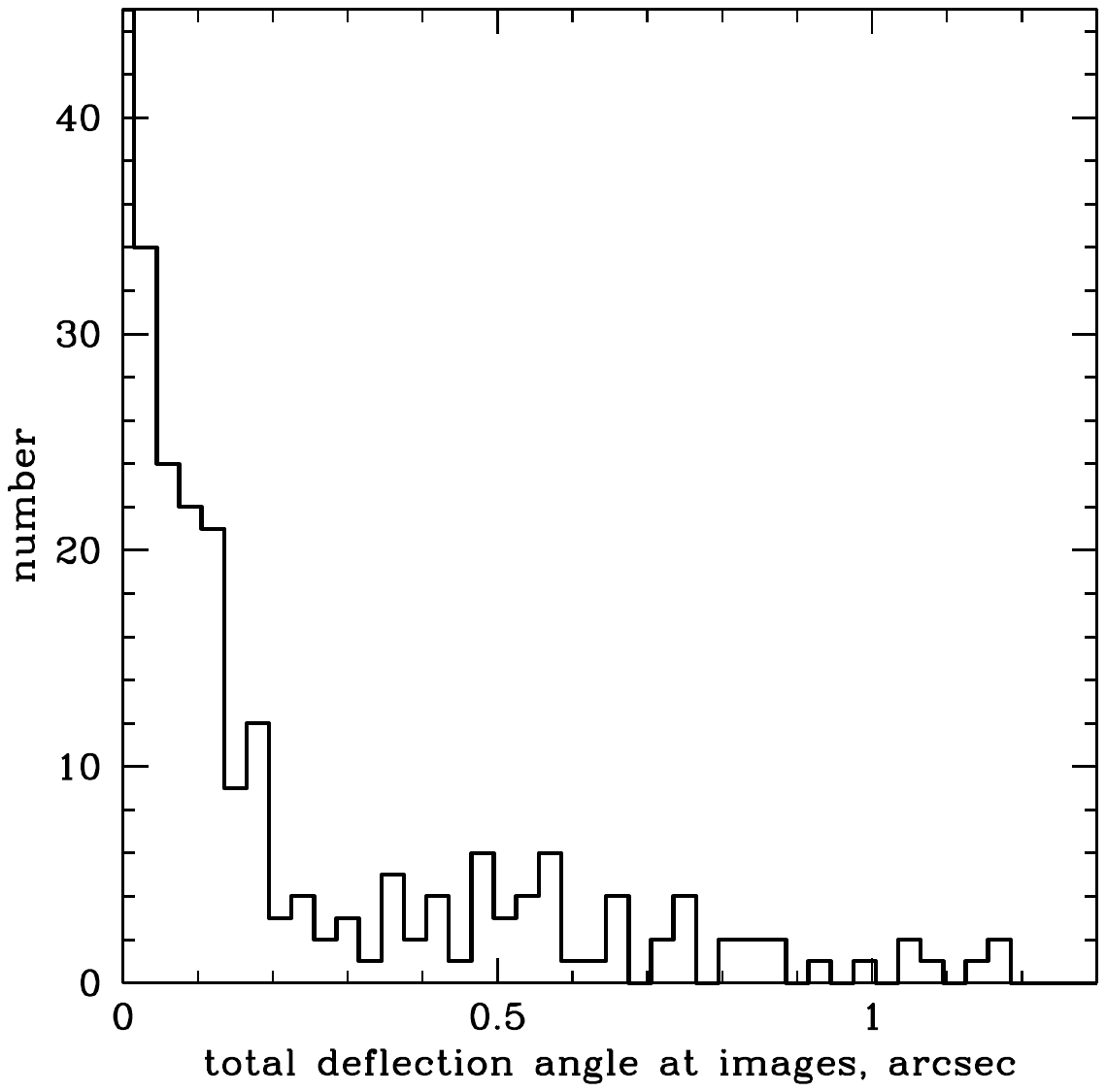}  
    \includegraphics[trim={0.7cm 5cm 1cm 5cm},clip,width=0.32\linewidth]{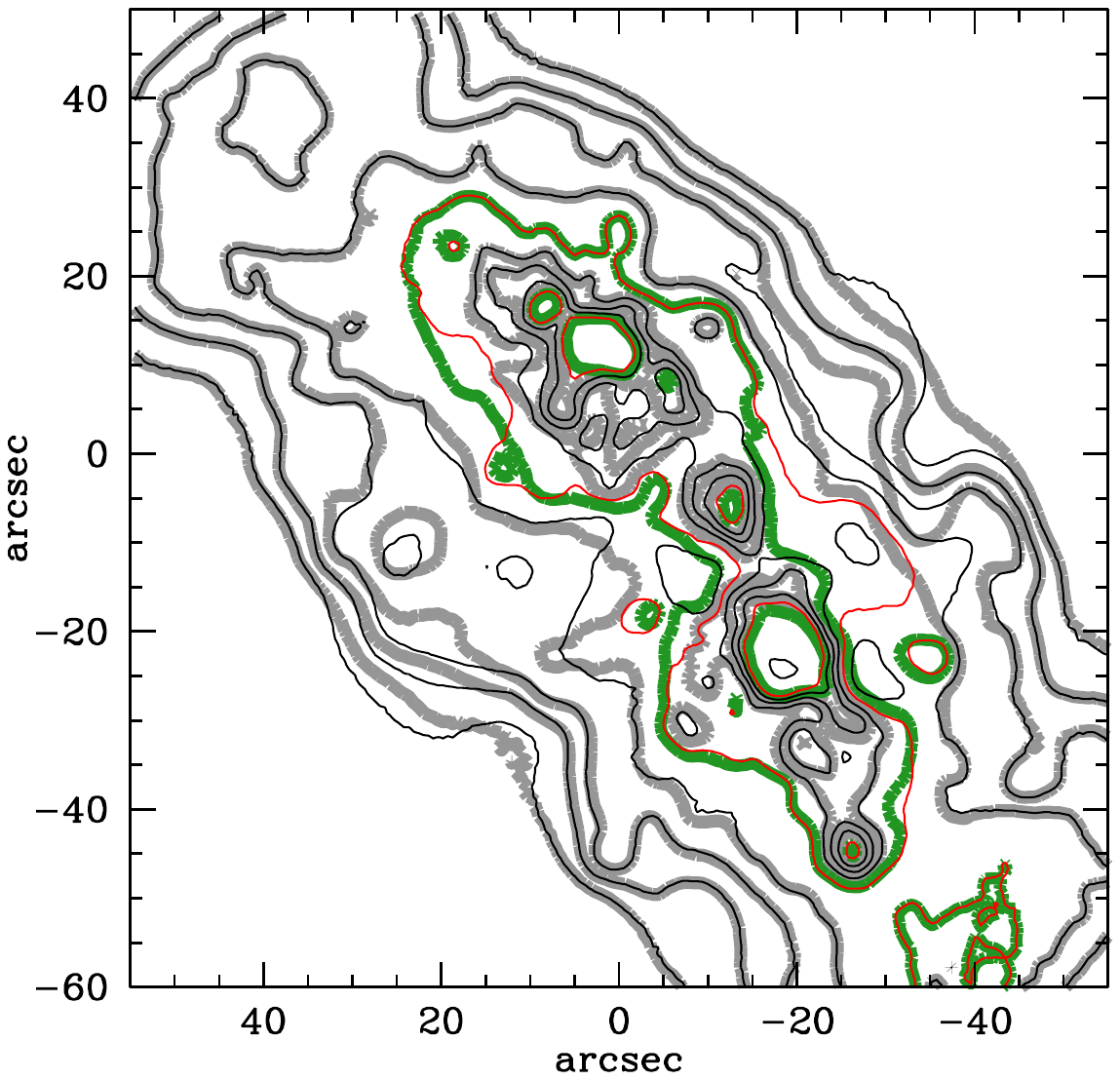} 
    \caption{Similar to Figure~\ref{fig:17A}, but for a different realization of ShaDes+.
    }
    \label{fig:17F}
\end{figure*}

In Figures~\ref{fig:17A} and \ref{fig:17F} we show two examples of ShaDes+ generated using linear algebra. We used 60 randomly placed points. Because this is significantly fewer than the number of images, the scale of resulting density perturbations is larger, as seen in the left panels. The median percent difference in these examples is 4.05\% and 2.80\%, respectively; see Table~\ref{tab:summary}. These values are larger than the other entries in the Table that correspond to ShaDes, i.e., where image positions are preserved exactly. MPD values of $\sim\,$few\% are more comparable to the typical MPD values found in \cite{Perera2025}, $\sim9\%$, who looked at differences between published models of MACS J0416. The $\Delta\kappa$ maps presented in that paper are examples of ShaDes+.


The black dots in the left panels of in Figures~\ref{fig:17A} and \ref{fig:17F} show positions of the observed images. Since the deflection angles at these positions are now non-zero (their distributions are shown in the middle panels), the magenta crosses show the positions after applying the deflection angles. The maximum deflections are about an arcsecond, and are comparable to those of typical lens reconstructions. The insets in the lower left zoom into the Warhol region of size $5''\times 5''$, where the offsets between black dots and magenta crosses are more visible. 

The right panels show {\tt Grale}'s $\kappa$ map of MACS J0416 as gray and green thick contours. When ShaDes+ are added the contours do not change dramatically; see black and red thin contours. The differences in the two sets of contours are comparable to, or smaller than those between various existing models. 

Since ShaDes+ are more representative of the actual differences between various models of the same cluster, we calculate what effect these have on the estimation of $H_0$. The light and dark green histograms in Figure~\ref{fig:H0} show the fractional difference in derived $H_0$ when using {\tt Grale} ($\mathcal{M}(\bm{\theta})$), and {\tt Grale} with ShaDes ($\mathcal{S}(\bm{\theta})$). Typical values are somewhat below $1\%$, and maximum differences reach $10\%$. These values can be compared to those presented in \cite{kelly2023} for Supernova Refsdal in MACS J1149: the fractional difference between the two best models is $2\%$, while the dispersion between all 8 models is about $10\%$. This supports the idea that the approximate ShaDes+ rather than the exact ShaDes are the relevant transformations.

\new{\cite{suyu2026} report observed time delays and corresponding lens model predictions for supernova Encore in galaxy cluster MACS J0138-2155, which hosts 23 images, including multiple images of two supernovae from the same galaxy, Requiem and Encore. The observed delay between Requiem images 2a and 2b is around 125 days, whereas all lens models presented in that paper have recovered time delay values around 50 days, and most disagree with observations by several $\sigma$ (see their Fig.4). For Requiem images 2a and 2c the observed time delay is also around 125 days, while the recovered value hover around 75 days, and some models are within $<1\sigma$ of the data. These differences between lens-model recovered and observed time delays are significantly larger than what our ShaDes predict in the case of clusters with hundreds of images, like MACS J0416 and MACS J1149. MACS J0138 does not have nearly as many multiple images, and hence constraints on its mass distribution, therefore it is likely that clusters with smaller number of images are less constrained.}

\section{ShaDes and other degeneracies}\label{sec:disc}

As newly characterized lensing degeneracies, it is interesting to consider how ShaDes and ShaDes+ fit into the existing framework of degeneracies.

ShaDes can be compared to another type of degenerate solutions, examined in \cite{liesenborgs2024}, in the context of galaxy cluster SDSS J1004+4112, and its dark mass clump. In that paper the lensing potential in the neighborhood of a (possibly extended) image was preserved, ensuring that all image properties, their positions, magnifications and time delays stay the same after the transformation. This is in contrast to ShaDes, where only the deflections exactly at the (point) image positions are unchanged. This implies that ShaDes preserve image positions only; other image properties, namely magnifications and time delays will be different compared to the original mass model $\mathcal{M}(\bm{\theta})$. 

%How do ShaDes fit into the broader set of lensing degeneracies? There are two ways of answering this question. On one hand, we can define ShaDes strictly as degeneracies constructed using the linear algebra procedure we used in this paper. In that case,  ShaDes are a new class of degeneracies, alongside the known ones, and can be compared to the latter.

Since ShaDes are defined as mass transformations that do not change deflection angles at the images, they are different from the mass sheet degeneracy \citep{falco1985,saha2000}. ShaDes+ are also different from MSD, as the two apply qualitatively different types of displacements to the sources. ShaDes and ShaDes+ have some overlap with the monopole degeneracy \citep{saha2000,liesenborgs2008b,liesenborgs2024} because some fraction of their $\Delta\kappa$ can be ascribed to circular redistribution of mass density between images. However, ShaDes and ShaDes+ are more general as they can have non-zero $\Delta\kappa$ at the locations of images. 

ShaDes+ share a common property with the generalized MSD studied in \cite{schneider2014b,schneider2019} and \cite{teodori2026} in the sense that all of these introduce additional smooth lensing potentials at a range of source redshifts.

ShaDes and ShaDes+ share a different property with the generalized MSD described in \cite{liesenborgs2008a}, the monopole degeneracy, and the SPT \citep{schneider2014a,unruh2017}. All these introduce density perturbations that have fluctuations on spatial scales smaller, and sometimes significantly smaller than the extent of the main lens. Neither the original MSD nor its multi-plane generalization do that. This distinction based on the scale of the perturbations is important when assessing the existence of mass substructures in galaxy cluster. The original and multi-plane MSD cannot create or hide substructure, while ShaDes, ShaDes+ and other degeneracies can.

%%There is another way of looking at shape degeneracies that would make them the superset of all known degeneracies. For example, MSD becomes a shape degeneracy if a ShaDe is a single flat mass sheet combined with a rescaled version of an existing mass model, and the sources displaced appropriately. While the current way we generate ShaDes is not optimal for reproducing MSD, this is merely an implementation problem, not a conceptual one. Continuing with this framework, if all the sources were at the same redshift, that would affect time delays by any amount one chooses, just like a standard MSD would. Similarly, monopole degeneracy is a subset of ShaDes, and so is SPT, as sources can be moved from their original positions under ShaDes.

\section{Conclusions}\label{sec:conc}

Lensing degeneracies are a nuisance for science goals addressed by galaxy clusters, as they are the source of systematic uncertainty. 
The range of degeneracies is wider than one would like.  Here we introduced and studied a new class of lensing degeneracies, shape degeneracies, or ShaDes, which do not have an analytic definition, but are pervasive in galaxy-scale and cluster-scale lenses.  As the name implies, these are mass perturbations that can take on any shape in the lens plane. These degeneracies leave image positions exactly the same as in the original model, or the same as the observed images, depending on what they are applied to, and either preserve the deflection angles and source positions, or allow them to be offset by a constant amount within any given source plane. Image positions are always preserved. Other image properties, like magnifications and time delays do change.

\neww{Time delays and hence the derived value of $H_0$ will typically be affected at the $1\%$ level, though can reach $10\%$ in some cases. Image magnifications are very important when clusters are used as cosmic telescopes to study high redshift sources. Though we did not consider changes in magnifications of background sources in this paper, given their important we will return to this in a later study.}

Our goal in this paper is to see how much difference between mass models can be introduced by, and hence attributed to these degeneracies without changing the image positions. Section~\ref{sec:macs0416} applied ShaDes to cluster lens MACS J0416, and concluded that typical fractional differences between cluster projected density distributions of two models related by ShaDes are $<1\!\%$ (see the last column, labeled MPD in Table~\ref{tab:summary}). 

\neww{One of the main motivations for studying ShaDes was to determine if dark mass clumps detected by free-form lens inversions models, but not detected by parametric methods could be due to different techniques arriving at shape-degenerate mass models. We showed that this is, in fact, possible, at least in the case of clump M2 in MACS J0416, if one takes in to account some differences between how well nearby images are reproduced in various models.}

The $<1\%$ difference mentioned above is significantly smaller than typical fractional differences between models from the literature, about $\sim9\%$. The reason is most likely that no two lens models reproduce image positions exactly, and ShaDes do not take that into account. We therefore introduced ShaDes+ which allow for image positions reconstructed by two different models to differ by amounts documented in the literature.

We conclude that, as with other lensing degeneracies it is not the exact ones that are relevant in the real world of modeling, but the approximate ones. In the case of shape degeneracies it means that ShaDes, while more interesting from the formal point of view, are less prevalent than ShaDes+ that allow images to be displaced by sub-arcsecond amounts. This conclusion is supported by the ShaDes+ estimate of how the derived value of $H_0$ will differ when calculated based on two models of the same cluster. Typical values differ by order of $1\%$, while in some rare cases the differences can be as large as $10\%$. These are consistent with what is seen for Supernova Refsdal in MACS J1149 \citep{kelly2023}, which has a similar number of images as MACS J0416 studied here.

A noteworthy difference between ShaDes and ShaDes+ concerns the sizes of density perturbation regions in the lens plane. In the case of exact degeneracies, ShaDes, the regions of positive and negative density perturbations are comparable in size to the typical separation between images. For approximate degeneracies, ShaDes+, these regions can be significantly larger than typical image separation, which is also more inline with what is seen when comparing different lens models.

Another important conclusion from our study is that tight groupings of images, like Warhol's 12 images in MACS J0416, are very effective at suppressing shape degeneracies. This suggests that mass reconstructions in the vicinity of such image configurations are more reliable than in other portions of clusters. Extended images, arising from extended sources \citep{acebron2024,schuldt2026} would also tend to suppress shape degeneracies. ShaDes+, which are an approximate degeneracy, add a small caveat to these statements. Based on our analysis, if positions of Warhol images are even slightly, $\sim\!0.2''$ different between two models (or a model and observed positions), then the projected density at Warhol's position can be different by $10\%$ between models. Whether this is significant or not depends on the science question being addressed.

Our analysis suggests that $\sim250$ images per cluster still leaves room for shape degeneracies, but doubling or tripling that number will lead to more accurate and precise cluster mass reconstructions. This is in agreement with the ``common wisdom" in the lensing community, as well as the findings of \cite{ghosh2020} who used synthetic clusters with up to a 1000 images. \new{The record holding cluster in terms of the number of images is MACS J0416, with a total of 415 multiple images, though only 303 of these are reliable \citep{rihtarsic2025}. Image numbers as large as 1000 may or may not be attainable with JWST at the limiting magnitude of 30AB. In that case, ShaDes, as well as ShaDes+ may remain a challenge for lens modeling. Cluster regions that have tight groupings of images, like images of knots from the same source galaxy, and extended arcs, will be less susceptible to shape degeneracies, and provide more accurate and precise reconstructions of the underlying mass distribution.}

\begin{acknowledgments}
LLRW would like to thank Prasenjit Saha for illuminating and fun discussions.
ML acknowledges the Centre National de la Recherche
Scientifique (CNRS) and the Centre National des
Etudes Spatiale (CNES) for support.
\end{acknowledgments}

\bibliographystyle{aasjournal}

\end{document}